\documentclass[journal]{IEEEtran}

\usepackage{dsfont}
\usepackage{amsmath,amsfonts,amssymb,amsthm}
\usepackage{graphicx,caption}
\usepackage[top=1in, bottom=1in, left=0.5in, right=0.5in]{geometry}
\usepackage{bm}
\newcommand*{\xhat}[1]{#1\kern-0.35em\hat{\phantom{#1}}}
\usepackage{mathrsfs,multicol,subfig}
\usepackage{cite}
\usepackage{amstext}
\usepackage{nidanfloat}
\usepackage{float}
\usepackage{nomencl}
\newcommand{\T}{\ensuremath{^\mathsf{T}}}

\newtheorem{theorem}{Theorem}[section]
\newtheorem{remark}{Remark}[section]
\newtheorem{lemma}{Lemma}[section]

\usepackage{hyperref}    
 \hypersetup{
     pdfstartview={FitH}, breaklinks=true,
    colorlinks=true, linkcolor=black,
     citecolor=black, filecolor=black,
     urlcolor=black,
 }

\title{Integrating Structure, Information Architecture and Control Design: Application to Tensegrity Systems}

\author{Raman~Goyal$^{1}$,
Manoranjan~Majji$^{2}$,
and~Robert~E.~Skelton$^{3}$
\thanks{$^{1}$TEES Postdoctoral Research Associate, Department of Aerospace Engineering, Texas A\&M University, College Station, TX 77843. Email: \textit{ramaniitrgoyal92@tamu.edu}}
\thanks{$^{2}$Assistant Professor, Department of Aerospace Engineering, Texas A\&M University, College Station, TX 77843. Email: \textit{mmajji@tamu.edu}}
\thanks{$^{3}$Professor, Wofford Cain Chair III, Department of Aerospace Engineering, Texas A\&M University, College Station, TX 77843. Email: \textit{bobskelton@tamu.edu}}}

\begin{document}

\maketitle

\begin{abstract}
A novel unified approach to jointly optimize structural design parameters, actuator and sensor precision and controller parameters is presented in this paper. The joint optimization problem is posed as a covariance control problem, where feasibility is achieved by bounding the covariance of the output as well as that of the control signals. The formulation is used to design a tensegrity system, where the initial prestress parameters, sensor and actuator precisions, and the control law are jointly optimized. Tensegrity system dynamics models linearized about an equilibrium point are used for system design, where minimality is ensured by constraint projection. The feedback loop is assumed to have a full-order dynamic compensator with its characteristic matrices chosen as optimization variables. The suboptimal solution of this non-convex system design problem is found by iterating over an approximated convex problem through the use of a convexifying potential function that enables the convergence to a stationary point. It is shown that for a linear dynamical system, the approximated joint optimization problem can be formulated using Linear Matrix Inequalities (LMIs).
\end{abstract}


\IEEEpeerreviewmaketitle

\section{Introduction}
\IEEEPARstart{T}{raditionally}, structure design and control design have been treated as two separate problems.
This follows the popular practice of dynamical system modeling followed by a control system design.  
This sequential design process is not optimal. Clearly, the control system design process can be optimized if the structure parameters can be adjusted to suit that choice. The idea is to design the structure and control algorithm to complement each other in achieving the required performance \cite{Goyal_ACC_2019}. 
Adhoc choices associated with the location, resolution and precision of the sensors and actuators in the structural system design process also contribute to the sub-optimal performance in the closed-loop operation of the dynamical system. 
An important element of overall system design includes an optimal selection of the structural parameters, controller design (control algorithm), along with the elements of the so-called information architecture (sensor and actuator precision and placement)\cite{IA_Paper_2008}.
% Similarly, controller design (control algorithm) and signal processing (location and precision of actuators and sensors) problems should be integrated to determine the required precision of sensors and actuators to guarantee the desired performance. 
Basically, a system-level design approach is required to make the best of all three disciplines where all components of the system are cooperatively designed to yield a specified system performance \cite{Goyal_ACC_2019}. 
Li \textit{et al.} \cite{IA_Paper_2008} integrated control design and selection of information architecture (actuator and sensor precision) to meet specified performance requirement (output covariance upper bound) formulating the constraints in Linear Matrix Inequalities (LMIs) \cite{Skelton_LMI_1998}. The integration of information architecture and control design is shown to be a convex problem for a linear time-invariant (LTI) plant with full-state feedback or full-order output feedback. Li \textit{et al.} \cite{IA_Paper_2008} also provides an ad hoc algorithm to obtain a minimal set of sensors or actuators. This is accomplished by repeatedly deleting the sensors or actuators with the least precision required until the design requirements cannot be met. Saraf \textit{et al.} \cite{Saraf_Bhatt_2017} made advancements in the information architecture theory by adding model uncertainty. Various researchers have also looked at the problem of finding a smaller set of sensors from a larger admissible set to obtain the desired closed-loop performance \cite{Oliveira_2000_Selection,Jager_2005, Balas_1999_sensor_selection}. \\

A framework to solve the instrument and control design problem is considered in \cite{Lu_Skelton_1999} and the problem of integrated structure and control design assuming known precisions of sensor and actuator is discussed in \cite{Lu_Skelton_2000_Integrating, Grandhi_1989_Optimization, Grigoriadis_1998_Integrated}. 
Grigoriadis \textit{et al.} \cite{Grigoriadis_1998_Integrated} provide a two-step solution for simultaneous design of structure and control by iterating over two convex sub-problems. First, the structure parameters are fixed and a controller is designed to satisfy the specified output covariance upper bound. In the second step, both the structure parameters and the controller are optimized such that the state covariance from the previous step is preserved. The algorithm iterates between these two steps to converge to a local minimum.
Lu \textit{et al.} \cite{Lu_Skelton_2000_Integrating} consider a more general structure parameterization and use mixed $H_2/H_{\infty}$ performance criteria, but still constrain the closed-loop state covariance matrix to be preserved in the second step. 
These approaches find a solution in reduced domain space, which may not necessarily be an optimal solution to the combined problem. Another approach to solving the integrated problem is by convexifying LMIs methods \cite{Camino_2000, Convexifying_Paper_2003}. In this approach, the authors first formulate a nonlinear matrix inequality to satisfy the performance requirement and then add another nonlinear matrix inequality to finally generate an LMI. There are some conditions on the added nonlinear matrix inequality (convexifying potential function) that guarantee that the solution will reach a stationary point \cite{CAMINO_2002_IFAC}. The contribution of this paper is to integrate information architecture and controller design along with the structure parameter as a free design variable. 
% The contribution of this paper is to add prestress in the tensegrity structures as another dimension to this system design approach by integrating structure parameters as an optimization variable along with information architecture and controller design.

Tensegrity structures are well suited to integrate structure and control design. 
As tensegrity provides good efficiency for both structural and control bases, it would be an ideal choice for various adaptive structures where structure and control parameters should be optimized simultaneously \cite{Bel_Smith_2010,IanSmith_CivilControl,Jager_2005}. Tensegrity structures are networks of axially loaded compressive (bars or struts) or tensile (strings or cables) members \cite{Snelson_1965,Ingber_1998}. A ``class-k" tensegrity structure has a maximum of $k$ bars connected with ball joints, restricting any torque/moment transfer between the elements. A ``class-1" structure is defined to have all the compressive members floating with no two compression members touching each other \cite{Skelton_2009_Tensegrity_Book}. 
% Accurate models of axially loaded members in tensegrity structures allow for easier and accurate control \cite{Goyal_Dynamics_2019,Djouadi_1998}.
An accurate system dynamics model due to only 1-dimensional deformation of strings, even with negative stiffness, allows for easier and accurate control \cite{Varol_2020_NegStiff,Goyal_Dynamics_2019,Djouadi_1998}.
The minimal mass architecture along with compliance and morphable shape characteristics makes tensegrity systems suitable for applications like planetary landers \cite{kim2020rolling, Goyal_2019_Buckling,Caluwaerts_2014}, deployable space structures \cite{Tibert_Pellegrino_2003_Mast,yang2019deployment,Sultan_Skelton_deployment,LandolfBarbarigos_Thesis}, flexible robots \cite{Sabelhaus_2017_ACC,Karnan_Goyal_2017_IROS,Paul_2006} and biomimetic robots \cite{sun2019adaptive,blissIwasakiCPG,mirletz2015goal}. 
Tensegrity structures are also robust to various kinds of loading conditions as the shape of the structure can be changed without changing the stiffness and one can also change the stiffness of a tensegrity structure without changing the shape \cite{zappetti2020phase,Varol_2020_NegStiff}.
The variable stiffness actuation and the easy addition of redundant strings enable robust design approaches making the structure well suited for human-robot interaction \cite{SoftRobots_IJRR,Varol_sensor_CST}. This redundancy increases the sensing requirement making the systems less reliable which dictates the need for only necessary and sufficient sensor placements \cite{Varol_sensor_CST}.
Aloui \textit{et. al.} integrate form-finding and damage detection in tensegrity structure \cite{Aloui_damage} and provide a framework for damage detection, sensor placement, and structural identification for tensegrity structures \cite{aloui2019theoretical}. Another research work for sensor/actuator placement through cellular morphogenesis of tensegrity structures has been proposed \cite{alouisensor}.

Other data-based and model-based control techniques have also been used for the control of tensegrity structures \cite{IanSmith_RL,Zhang_Levine_2017_ICRA,Wang_2020_RAL,IanSmith_ActMultObj,Koizumi_2012_ICRA}. The Central Pattern Generator (CPG) controller mimics the periodic system in biological neural circuits and has been used for controlling a tensegrity swimmer \cite{blissIwasakiCPG,blissIwasakiCPGexperimental}. Model Predictive Control (MPC) approaches have recently been used to control tensegrities \cite{PENG_2020_MPC,Sabelhaus_2020_CST}. 
Yang and Sultan \cite{Yang_Sultan_2017_IJRNC} designed an adaptive controller for these underactuated systems and further used robust control techniques for the controlled deployment and stabilization of a tensegrity membrane system using $\mathcal{H}_\infty$ control theory. A shape controller using output feedback linear parameter varying formulation is developed in \cite{Yang_Sultan_2017_LPVControl}. 
A pose estimation strategy was recently developed to control the position of the end effector of a tensegrity manipulator \cite{Varol_Manpltr} and a framework for automatic robotic assembly of tensegrity structures by controlling the motion of a specific end-effector is also recently proposed where the end-effector was attached to a standard industrial robot \cite{Varol_RoboPlan}. 

In this paper, an integrated domain formulation is developed to solve for all three design and decision variables i.e. control design, information architecture, and structure design simultaneously as an extension to the authors' previous work \cite{Goyal_ACC_2019}. The formulation is developed for a linear time-invariant system with structure parameters appearing linearly in the system matrices. A budget constraint is also considered to put an upper bound on the sensor/actuator precision with the cost of actuators and sensors assumed to be directly proportional to their precision value. 
The paper does not assume any constraint on the order of the controller but a full-order controller is used for the tensegrity example results.
Although the paper shows its application for tensegrity systems, the developed optimization framework applies to any LTI system with affine structure parameters. \\

% \textbf{Scope and limitation}
% In this research work, we use the tensegrity paradigm to achieve this by framing it as a covariance control problem. 

The paper is organized as follows: First, we describe a linear system in descriptor state-space form with plant matrices affine in structure parameters. The inverse of the noise of actuators and sensors are defined to be the precision of the actuators and sensors, respectively.
% The precision of the noisy actuators and sensors is considered as an optimization variable, and the cost of actuators and sensors is assumed to be directly proportional to their precision to add a budget constraint to the problem.
Then, we write matrix inequalities to stabilize the system and to satisfy output and control covariance constraints \cite{Skelton_LMI_1998}. The simultaneous design of a structure, information architecture, and controller results in nonlinear matrix inequalities even for linear systems. As the domain set is defined by nonlinear inequalities (possibly non-convex), a convexifying LMI method is used to solve this system design problem by approximating it to a convex problem \cite{Convexifying_Paper_2003}. The convexification is achieved by adding a nonlinear matrix inequality with certain conditions. The iteration on the approximated convex sub-problem guarantees the solution to reach a stationary point.
Second, an analytical formulation to linearize the nonlinear systems about an equilibrium point is provided for both full-order and reduced-order tensegrity dynamics models. Then, a novel algorithm to remove the modes corresponding to the bar length change is provided which allows for the minimal-order representation of the system. The robust control theory to reject the disturbances with different error bounds for linear systems is applied on a tensegrity robotic arm.
For the tensegrity system design, the linearized tensegrity dynamics model with initial prestress is used as the free structure parameter with force density in the strings to be the control input for the system.
The performance is defined to bound the displacement of some nodes while measuring the length of the strings or the position of the nodes.
The final output of the optimization problem would be the initial prestress (free structure parameter), the precision of sensors/actuators, and the characteristics matrices for the dynamic controller. A tensegrity beam example with 8 bars and 12 strings is used to demonstrate the utility of the joint optimization with different trade-off studies.

%%%%%%%%%%%%%%%%%%%%%%%%%%%%%%%%%%%%%%%%%%%%%%%%%%%%%%%%%%%%%%%%%%%%%%%%%%%%%%%%%%%%%%%%%%%%%%%%%%%%%%%%%%%%%%%%%

\section{Integrating Structure, Information Architecture and Control Design} \label{s:IASD}
This section presents the technical formulation to include information precision and architecture in control design along with the structure design. This enlarges the set of solved linear control problems, from solutions of linear controllers with prespecified sensors/actuators and structure parameters, to solutions which specify the sensor/actuator requirements jointly with the structure parameters and the control solution \cite{IA_Paper_2008}. Some parts of this section are presented in our recent paper \cite{Goyal_ACC_2019}.

A continuous linear time-invariant system is described by the following \textit{descriptor} state-space representation:

\begin{align}
E(\alpha)\dot{x} &= A(\alpha)x +Bu + D_p(\alpha)w_p +D_a(\alpha)w_a,\label{state_eqn2} \\
y &= C_y(\alpha)x, \hspace{0.5 pc} \text{(output)} \label{output_eqn2} \\ 
z &= C_zx+D_sw_s, \hspace{0.5 pc} \text{(measurement)} \label{meas_eqn2}
\end{align}

\noindent where $x \in \mathbb{R}^n$ is the state of the system, $u \in \mathbb{R}^m$ is the control vector, $y \in \mathbb{R}^p$ is the output of the system, $z \in \mathbb{R}^\ell$ denotes the measurement vector, and $w_i$ for $i = {a,s,p}$ are noisy inputs to the system. The vector $\alpha$ consists of generalized variable structure parameters that can be treated as decision variables in the system design problem. 
% For example, in a second-order system, $\alpha$ would make up various mass, stiffness, and damping parameters. 
It is assumed that the following matrices are affine in the parameters $\alpha$: $A(\alpha),E(\alpha), D_p(\alpha), D_a(\alpha)$ and $C_y(\alpha)$.
Note that in a typical second-order system, it is useful to adopt the descriptor representation in order to preserve the affine property of the system mass matrix. The matrix $E(\alpha)$ is also assumed to be full rank. 

In the above model Eqs.~(\ref{state_eqn2}-\ref{meas_eqn2}), the actuator noise is defined by $w_a$, sensor noise by $w_s$, and ambient process noise by $w_p$. These vectors are modeled as independent zero mean white noises with intensities $W_a$, $W_s$ and $W_p$, respectively, i.e.:

\begin{gather}
\mathbb{E}_{\infty}(w_i) = 0, \\
\mathbb{E}_{\infty}(w_iw_i\T) = W_i \delta(t-\tau),
\end{gather}
where $i = {a,s,p}$, and $\mathbb{E}_{\infty}(x) = \lim_{t \rightarrow \infty} \mathbb{E}(x)$ which denotes the asymptotic expected value of the random variable $x$. We assume the process noise intensity $W_p$ to be known and fixed. The actuator and sensor precisions are defined to be inversely proportional to the respective noise intensities.

\begin{gather}
\Gamma_a \triangleq W_a^{-1}, ~~
\Gamma_s \triangleq W_s^{-1}.
\end{gather}
We also define the vectors $\gamma_a$ and $\gamma_s$ such that:
\begin{gather}
\Gamma_a \triangleq \text{diag}(\gamma_a), ~~
\Gamma_s \triangleq \text{diag}(\gamma_s).
\end{gather}

As defined in \cite{IA_Paper_2008}, we associate a price to each actuator/sensor that is inversely proportional to the noise intensity associated with that instrument. In this work, prices associated with the structure parameters are also considered. Therefore, the total design price can be expressed as:
\begin{equation}
\$ = p_a\T\gamma_a+p_s\T\gamma_s+p_{\alpha}\T\alpha,
\end{equation}
where $p_a$, $p_s$ and $p_{\alpha}$ are vectors containing the price per unit of actuator precision, sensor precision and price per unit of structure parameter, respectively. 

Now, the problem to be solved is defined as to design a dynamic compensator of the form:
\begin{equation}
\begin{split}
\dot{x}_c = A_cx_c+B_cz, \\
u = C_cx_c+D_cz,
\end{split} \label{dyn_comp_control}
\end{equation}
and simultaneously select the structure parameter values, appropriate actuator and sensor precisions such that the following constraints are satisfied:

\begin{equation}
\begin{split}
\$ < \bar{\$},~~\gamma_a < \bar{\gamma}_a ,~~\gamma_s < \bar{\gamma}_s, ~~\bar{\alpha}_L < \alpha < \bar{\alpha}_U, \\
~~~\mathbb{E}_{\infty}(uu\T) < \bar{U}, ~~
\mathbb{E}_{\infty}(yy\T) < \bar{Y} 
\label{constraints2}
\end{split}
\end{equation} 
for given $\bar{\$}$, $\bar{U}$,  $\bar{Y}$, $\bar{\gamma}_a$, $\bar{\gamma}_s$, $\bar{\alpha}_L$, and $\bar{\alpha}_U$. 

\subsection{Solution to Dynamic Compensation Problem}

\begin{theorem} \label{dyn_comp_thm}
    Let a continuous time-invariant linear system be described by the descriptor state space equation (\ref{state_eqn2}),
    % where $(E^{-1}(\alpha)A(\alpha),E^{-1}(\alpha)B)$ is a controllable pair
    the output equation (\ref{output_eqn2}) and the measurement equation (\ref{meas_eqn2}). There exist controller matrices $A_c$, $B_c$, $C_c$ and structure parameters $\alpha$ such that the cost and performance constraints (\ref{constraints2}) are satisfied, if and only if for some constant matrix $G$, there exists a symmetric matrix Q, vectors $\gamma_a$, $\gamma_s$ and $\alpha$
    such that the following LMIs are satisfied:
    
    \begin{gather}
    p_a\T\gamma_a+p_s\T\gamma_s+p_{\alpha}\T\alpha < \bar{\$},  \label{theorem_eqn21} \\
    \gamma_a < \bar{\gamma}_a,~~
    \gamma_s < \bar{\gamma}_s, \label{theorem_eqn23} \\
    \bar{\alpha}_L < \alpha < \bar{\alpha}_U,   \label{theorem_eqn24} \\
    \begin{bmatrix} \bar{U} & M_{cl} \\ M_{cl}\T & Q \end{bmatrix} > 0, \hspace{.5 pc}
    \begin{bmatrix} \bar{Y} & C_{cl} \\ C_{cl}\T & Q \end{bmatrix} > 0, \label{theorem_eqn26} \\
    \begin{bmatrix} (\star) & B_{cl} & A_{cl} & E_{cl} \\ B_{cl}\T & -W^{-1} & 0 & 0 \\ A_{cl}\T & 0 & -Q & 0 \\ E_{cl}\T & 0 & 0 & -Q \end{bmatrix} < 0, \label{theorem_eqn27} 
    \end{gather}
    where 
    \begin{gather*}
    (\star) = -(A_{cl}-E_{cl})G\T-G(A_{cl}-E_{cl})\T+GQG\T,\\
    W = \begin{bmatrix} W_p & 0 & 0\\ 0 & W_a & 0 \\ 0 & 0 & W_s \end{bmatrix}, \hspace{1 pc}
    A_{cl} = \begin{bmatrix} A(\alpha)  & BC_c \\ B_cC_z & A_c \end{bmatrix},\\ 
    E_{cl} = \begin{bmatrix} E(\alpha) & 0 \\ 0 & I_n \end{bmatrix},
    B_{cl} = \begin{bmatrix} D_p(\alpha) & D_a(\alpha) & 0 \\ 0 & 0 & B_cD_s \end{bmatrix}, \\
    C_{cl} = \begin{bmatrix} C_y(\alpha) & 0 \end{bmatrix}, \hspace{1 pc}
    M_{cl} = \begin{bmatrix} 0 & C_c \end{bmatrix},
    \end{gather*}
    
    \noindent and $I_n$ is a $n \times n$ identity matrix. 
\end{theorem}

\noindent
\textbf{Proof.} Define the augmented vector $\tilde{x}$ and $w$ as:

\begin{equation}
\tilde{x}\T = \begin{bmatrix} x\T & x_c\T  \end{bmatrix}, \hspace{1 pc}
w\T = \begin{bmatrix} w_p\T & w_a\T & w_s\T \end{bmatrix}.
\end{equation}

The closed loop dynamics for the state $\tilde{x}$ can be written in the descriptor representation form along with the output and control equations as:
\begin{gather}
E_{cl}\dot{\tilde{x}} = A_{cl}\tilde{x}+B_{cl}w, \label{augmented_state} \\
y = C_{cl}\tilde{x}, \label{augmented_output} \\
u = M_{cl}\tilde{x}+F_{cl}w, \label{augmented_control}
\end{gather} 
    where all the close loop matrices can easily be obtained from the system equations (\ref{state_eqn2}-\ref{meas_eqn2}) and controller equations (\ref{dyn_comp_control}).
    
%%%%%%%%%%%%%%%%%%%%%%%%%%%%%%%%%%%%%%%%%%%%%%%%%%%%%%%%%%%%%%%%%%%%%%%%%%%%%%%%%%%%%%%%%%%%%
    % where
    % \begin{gather*}
    % W = \begin{bmatrix} W_p & 0 & 0\\ 0 & W_a & 0 \\ 0 & 0 & W_s \end{bmatrix} \hspace{1 pc}
    % E_{cl} = \begin{bmatrix} E(\alpha) & 0 \\ 0 & I_n \end{bmatrix}\\ 
    % A_{cl} = \begin{bmatrix} A(\alpha)+BD_cC_z & BC_c \\ B_cC_z & A_c \end{bmatrix} \hspace{.5 pc}
    % B_{cl} = \begin{bmatrix} D_p(\alpha) & D_a(\alpha) & BD_cD_s \\ 0 & 0 & B_cD_s \end{bmatrix} \\\hspace{1 pc}
    % C_{cl} = \begin{bmatrix} C_y(\alpha) & 0 \end{bmatrix} \hspace{1 pc}
    % M_{cl} = \begin{bmatrix} D_cC_z & C_c \end{bmatrix} \hspace{1 pc}
    % F_{cl} = \begin{bmatrix} 0 & 0 & D_cD_s \end{bmatrix}
    % \end{gather*}
    % and $I_n$ is the $n \times n$ identity matrix. 
%%%%%%%%%%%%%%%%%%%%%%%%%%%%%%%%%%%%%%%%%%%%%%%%%%%%%%%%%%%%%%%%%%%%%%%%%%%%%%%%%%%%%%%%%
Defining $\bar{A}_{cl} = E_{cl}^{-1}A_{cl}$ and $\bar{B}_{cl} = E_{cl}^{-1}B_{cl}$ and rearranging equation (\ref{augmented_state}) gives:

\begin{equation}
\dot{\tilde{x}} = \bar{A}_{cl}x+\bar{B}_{cl}w.
\end{equation}

It is a standard result that the above closed loop system is stable if and only if there exists a positive definite symmetric matrix $X$ such that:

\begin{equation} 
\bar{A}_{cl}X+X\bar{A}_{cl}\T+\bar{B}_{cl}W\bar{B}_{cl}\T < 0. \label{stable_ineq21}
\end{equation}

Multiplying the inequality (\ref{stable_ineq21}) from left by $E_{cl}$ and from right by $E_{cl}\T$ yields:

\begin{equation}
{A}_{cl}XE\T_{cl}+E_{cl}X{A}_{cl}\T+{B}_{cl}W{B}_{cl}\T < 0 \label{stable_ineq22}.
\end{equation}

Applying Schur's complement on (\ref{stable_ineq22}) gives: 

\begin{equation}
\begin{bmatrix} {A}_{cl}XE\T_{cl}+E_{cl}X{A}_{cl}\T & B_{cl} \\ B_{cl}\T & -W^{-1} \end{bmatrix} < 0. \label{stable_ineq23}  
\end{equation}

It can be shown that after substitution of $E_{cl}$, $A_{cl}$ and $B_{cl}$, inequality (\ref{stable_ineq23}) does not form an LMI since it is not affine in the decision variables $A_c$, $B_c$, $\alpha$, etc. On completing the squares, the inequality (\ref{stable_ineq23}) can be rewritten as:

\begin{align}
\begin{bmatrix} 
\begin{split}
&A_{cl}XA_{cl}\T+E_{cl}XE_{cl}\T\\&-(A_{cl}-E_{cl})X(A_{cl}-E_{cl})\T 
\end{split} & B_{cl} \\  & B_{cl}\T & -W^{-1} \end{bmatrix} <0. \label{stable_ineq24}
\end{align}

Defining $\delta \triangleq (A_c,B_c,C_c,\gamma_a,\gamma_s,\alpha,Q)$, $Q \triangleq X^{-1}$, and applying Schur's complement, we can write the inequality (\ref{stable_ineq24}) as:

\begin{equation}
\mathbb{F}(\delta) \triangleq \begin{bmatrix} (\bullet) & B_{cl} & A_{cl} & E_{cl} \\ B_{cl}\T & -W^{-1} & 0 & 0 \\ A_{cl}\T & 0 & -Q & 0 \\ E_{cl}\T & 0 & 0 & -Q \end{bmatrix}<0, \label{stable_ineq25}
\end{equation}

\noindent where $(\bullet) = -(A_{cl}-E_{cl})X(A_{cl}-E_{cl})\T$. Note that $\mathbb{F}(\delta)$ is not an LMI.
Let us introduce the convexifying algorithm Lemma to write a new LMI.

\begin{lemma} \label{convex_lemma}
    \noindent
    \textbf{Convexifying Algorithm Lemma.}
    Let $\delta$, $\eta$ belong to a convex set $\phi$, and $\mathbb{F}(\delta)$ be a first order differentiable non-convex matrix function. A convexifying potential function is a first order differentiable function $\mathbb{G}(\delta,\eta)$ such that the function $\mathbb{F}(\delta)+\mathbb{G}(\delta,\eta)$ is convex in $\delta$ for all $\delta,\eta \in \phi$. Thus, if $\mathbb{F}(\delta)$ satisfies certain conditions, a stationary point of the non-convex optimization problem    
    
    \begin{equation}
    \bar{\delta} = \arg \min_{\delta \in \Omega} f(\delta), \hspace{0.5 pc} \Omega = \{\delta \in \phi| \mathbb{F}(\delta)<0   \}, \label{lemma_eq1}
    \end{equation}
    
    \noindent can be obtained by iterating over a sequence of convex subproblems given by
    
    \begin{equation}
    \bar{\delta}_{k+1} = \arg \min_{\delta \in \Omega_k} f(\delta), \Omega_k = \{\delta \in \phi| \mathbb{F}(\delta)+\mathbb{G}(\delta,\delta_k)<0   \}. \label{lemma_eq2}
    \end{equation}
    
    To ensure that the optimality conditions of both optimization problems (\ref{lemma_eq1}) and (\ref{lemma_eq2}) are identical, the potential function $\mathbb{G}$ should be non-negative definite
    with $\mathbb{G}(\delta,\eta)=0$ if and only if $\delta=\eta$.
\end{lemma}

% \begin{remark}
\noindent The previous Lemma is proven and discussed in further detail in \cite{Convexifying_Paper_2003}. Although the Convexifying Algorithm will converge to a stationary point, a global solution is not guaranteed.
% \end{remark}

To use the previous Lemma, let us define the matrix $G$ as:
\begin{equation}
G(\eta) \triangleq (A_{cl}-E_{cl})X,
\end{equation}
and the convexifying potential function as:

\begin{align}
\mathbb{G}(\delta,\eta) \triangleq & \begin{bmatrix}
     (\ast) & \mathbf{0} \\ \mathbf{0} & \mathbf{0}\end{bmatrix} \geq  0, \\
 \nonumber    (\ast) = (A_{cl}-E_{cl}-G(\eta)Q)&X(A_{cl}-E_{cl}-G(\eta)Q)\T.
%  \\
% &\mathbb{G}(\delta,\eta) \geq  0.
\end{align}

The matrix function $\mathbb{F}(\delta)+\mathbb{G}(\delta,\eta)$:

\begin{equation}
\begin{bmatrix} (\star) & B_{cl} & A_{cl} & E_{cl} \\ B_{cl}\T & -W^{-1} & 0 & 0 \\ A_{cl}\T & 0 & -Q & 0 \\ E_{cl}\T & 0 & 0 & -Q \end{bmatrix} < 0, \label{stable_ineq26}
\end{equation}

\noindent where
$(\star) = -(A_{cl}-E_{cl})G\T-G(A_{cl}-E_{cl})\T+GQG\T,$ is convex, where the dependency of the matrix $G$ on $\eta$ is omitted for brevity. The function $\mathbb{G}(\delta,\eta)$ satisfies the convexifying assumptions since it is
positive semidefinite and $\mathbb{G}(\delta,\eta)=0$ if and only if $\delta = \eta$. Furthermore, using Lemma (\ref{convex_lemma}), it can be shown that any solution to (\ref{stable_ineq26}) will also satisfy (\ref{stable_ineq25}) \cite{Convexifying_Paper_2003}. 
% Also, note that inequalities (\ref{stable_ineq26}) and (\ref{theorem_eqn27}) are identical. 

The second constraint of the constraint set (\ref{constraints2}) can be evaluated by substituting in the expression for the control law (\ref{augmented_control}) given by:
\begin{align}
\mathbb{E}_{\infty}(uu\T) = \mathbb{E}_{\infty}((M_{cl}\tilde{x})&(M_{cl}\tilde{x})\T)+\mathbb{E}_{\infty}((F_{cl}\tilde{w})(F_{cl}\tilde{w})\T),\\
\mathbb{E}_{\infty}(uu\T)& < \bar{U}, \\
M_{cl}XM_{cl}\T+&F_{cl}WF_{cl}\T < \bar{U}.
\end{align}
The second term can grow unbounded if $F_{cl} \neq 0$. Hence, substituting for $F_{cl} = 0$ in the above equation gives:

\begin{equation}
M_{cl}XM_{cl}\T < \bar{U} \hspace{.5 pc} \text{and} \hspace{.5 pc} D_c = 0 ~ \text{($D_s$ is full rank)}.
\end{equation}
and applying Schur's complement to first term results in left inequality of equation (\ref{theorem_eqn26}). It is then straightforward to show that the last constraint of (\ref{constraints2}) is satisfied if and only if:

\begin{equation}
C_{cl}XC_{cl}\T < \bar{Y}.
\end{equation}
Applying Schur's complement to this inequality results in (\ref{theorem_eqn26}). 
Finally, first four constraints in (\ref{constraints2}) are first four inequalities of Theorem \ref{dyn_comp_thm}.
% , (\ref{theorem_eqn21}),  (\ref{theorem_eqn23}), and (\ref{theorem_eqn24}) 
% must be satisfied. 
\qed
 
\begin{remark}
    Assume that $\bar{\gamma}_a$ and $\bar{\gamma}_s$ are dictated by the marketplace. Let four parameters out of the set ($\bar{\alpha}_L$, $\bar{\alpha}_U$,$\bar{\$}$, $\bar{U}$, $\bar{Y}$) be hard constraints and let the fifth parameter, denoted $\bar{z}$, be any value for which the LMIs of Theorem \ref{dyn_comp_thm} are feasible. The following iterative algorithm takes advantage of Lemma \ref{convex_lemma} to find an extrema for $\bar{z}$ (a minimum if $\bar{z} = \bar{\alpha}_U, \bar{\$}, \bar{U}, \bar{Y}$ or a maximum if $\bar{z} = \bar{\alpha}_L$).    \\
\end{remark}

\noindent
\textbf{Extrema-Finding Algorithm using the Convexifying Potential Function}
\begin{itemize}
    \item Set fixed nominal values for $\bar{z}_0$ and $\alpha_0$. Compute controller matrices $A_{c,0}$, $B_{c,0}$, $C_{c,0}$, precision vectors $\gamma_{a,0}$, $\gamma_{s,0}$ and inverse covariance matrix $Q_0$ according to \cite{IA_Paper_2008} or some alternative method. Set $\epsilon$ to some prescribed tolerance and $k=0$.\\
    
    \item \textbf{Repeat:} Set $G_k \leftarrow (A_{cl}(\alpha_k)-E_{cl}(\alpha_k))Q^{-1}_k$ \newline -For fixed $G=G_k$, find the extrema of $\bar{z}$ for which the LMIs of Theorem \ref{dyn_comp_thm} are feasible \newline
    -Denote the solution ($\bar{z}_{k+1}$, $\alpha_{k+1}$, $A_{c,k+1}$, $B_{c,k+1}$, $C_{c,k+1}$,$\gamma_{a,k+1}$, $\gamma_{s,k+1}$,$Q_{k+1}$)
    % -Denote the solution ($\bar{z}^*$, $\alpha^*$, $A_{c}^*$, $B_{c}^*$, $C_{c}^*$,$\gamma_{a}^*$, $\gamma_{s}^*$,$Q^*$)
    % \newline
    % -Set ($\bar{z}_{k+1}$, $\alpha_{k+1}$, $A_{c,k+1}$, $B_{c,k+1}$, $C_{c,k+1}$,$\gamma_{a,k+1}$, $\gamma_{s,k+1}$,$Q_{k+1}$) $\leftarrow$ ($\bar{z}^*$, $\alpha^*$, $A_{c}^*$, $B_{c}^*$, $C_{c}^*$,$\gamma_{a}^*$, $\gamma_{s}^*$,$Q^*$)
    \newline -Set k = k+1 \\
    
    \item \textbf{Until:} $\| \bar{z}_{k}-\bar{z}_{k-1} \| < \epsilon$.\\
\end{itemize}

\noindent The above algorithm allows optimizing the structure design and information architecture along with the design of the dynamic output feedback controller. Notice that there is no constraint on the order of the controller in the formulation. More detailed discussion on the reduced-order controllers can be found in \cite{IA_Paper_2008}.

\section{State Feedback Control Problem}
The full-state feedback control problem considers the case where all the states are available for measurement with no noise. Thus the information architecture in this situation reduces to finding the precision and location of actuators only.

\subsection{Problem Statement}
Let us consider the situation where full-state feedback is available for measurement, i.e. $C_y$ is invertible. The system can now be described as:
% in the \textit{descriptor} state-space representation by:
\begin{gather}
E(\alpha)\dot{x} = A(\alpha)x +Bu + D_p(\alpha)w_p +D_a(\alpha)w_a, \label{state_eqn1} \\
y = C_y(\alpha)x, \hspace{0.5 pc} \text{(output)} \label{output_eqn1}
\end{gather}
As there is no measurement noise, the total design price is expressed as:
\begin{equation}
\$ = p_a\T\gamma_a+p_{\alpha}\T\alpha.
\end{equation}
The state feedback problem can now be defined as:

Design a state feedback controller $u = -Kx$ and simultaneously select the structure parameters and the actuator precisions such that the following constraints are satisfied:
\begin{equation}
\begin{split}
\$ < \bar{\$},\gamma_a < \bar{\gamma}_a,  \bar{\alpha}_L < \alpha < \bar{\alpha}_U,~
\mathbb{E}_{\infty}(uu\T) < \bar{U},
\mathbb{E}_{\infty}(yy\T) < \bar{Y} , \label{constraints1}
\end{split}
\end{equation} 
for given $\bar{\$}$, $\bar{U}$, $\bar{Y}$, $\bar{\gamma}_a$, $\bar{\alpha}_L$, and $\bar{\alpha}_U$. 

\subsection{Solution to State Feedback Control Problem}

\begin{lemma} \label{state_feedback_thm}
Let a continuous time-invariant linear system be described by the descriptor state space equation (\ref{state_eqn1})
% where $(E^{-1}(\alpha)A(\alpha),E^{-1}(\alpha)B)$ is a controllable pair 
and the output equation (\ref{output_eqn1}). There exists a controller gain $K$ and structure parameters $\alpha$ such that the cost and performance constraints (\ref{constraints1}) are satisfied if and only if for some constant matrix $G$ there exists a symmetric matrix Q and vectors $\gamma_a$ and $\alpha$ such that the following LMIs are satisfied:
\begin{gather}
p_a\T\gamma_a+p_{\alpha}\T\alpha < \bar{\$} , \label{theorem_eqn11} \\
\gamma_a < \bar{\gamma}_a,~~~
\bar{\alpha}_L < \alpha < \bar{\alpha}_U ,  \label{theorem_eqn13} \\
\begin{bmatrix} \bar{U} & K \\ K\T & Q \end{bmatrix} > 0 , \hspace{0.5 pc}
\begin{bmatrix} \bar{Y} & C_y(\alpha) \\ C_y(\alpha)\T & Q \end{bmatrix} > 0, \label{theorem_eqn15} \\
\begin{bmatrix} (\star) & B_{cl} & A_{cl} & E(\alpha) \\ B_{cl}\T & -W^{-1} & 0 & 0 \\ A_{cl}\T & 0 & -Q & 0 \\ E\T(\alpha) & 0 & 0 & -Q \end{bmatrix} < 0, \label{theorem_eqn16}
\end{gather}
where 
\begin{gather*}
(\star) = -(A_{cl}-E(\alpha))G\T-G(A_{cl}-E(\alpha))\T+GQG\T, \\
W = \begin{bmatrix} W_p & 0 \\ 0 & W_a \end{bmatrix}, \hspace{.5 pc}
A_{cl} = A(\alpha)-BK, \\
B_{cl} = \begin{bmatrix} D_p(\alpha) & D_a(\alpha) \end{bmatrix}.
\end{gather*}
\end{lemma}

\begin{proof}
 The proof is excluded as it follows very closely with Theorem \ref{dyn_comp_thm}. 
\end{proof}

\section{Linearized Tensegrity Dynamics}
The nonlinear dynamics of a tensegrity structure of any complexity is derived in \cite{Goyal_Dynamics_2019}. To apply the optimization formulation developed in Section \ref{s:IASD}, the nonlinear equations are linearized and represented in the descriptor form such that prestress appears as a linear free variable. This section also presents the reduced-order linear model for the class-k tensegrity structures.

\subsection{Class-1 Linearized Dynamics}\label{L:linear_class1}
\begin{lemma}
The linearized dynamics of any class-1 tensegrity system in terms of linear variation in nodal coordinates $\tilde{n}$ can be written as:
\begin{align}
    \mathcal{M}_1 \ddot{\tilde{n}} + \mathcal{D}_1 \dot{\tilde{n}} + \mathcal{K}_1 \tilde{n} = \mathcal{P}_1 \tilde{w}+\mathcal{B}_1 \tilde{\gamma}, \label{eq:lnr:cls1}
\end{align}
where $\tilde{\gamma}$ is the linear variation in the force density, $\tilde{w}$ is the linear variation in external force, and 
\begin{align}
    &\mathcal{M}_1 \triangleq \mathcal{T}\T M_{br}\mathcal{T}, ~ \mathcal{D}_1 \triangleq \mathcal{T}\T D_{br}\mathcal{T}, \\
    &\mathcal{K}_1 \triangleq \mathcal{T}\T K_{br}\mathcal{T}+\mathcal{P}_1 \hspace{0.2pc}(C_s\T \otimes I)\hspace{0.2pc} \widehat{(\bar{\gamma} \otimes \mathds{1})}(C_s \otimes I),
\end{align}
\begin{align}
    &\mathcal{P}_1 \triangleq \mathcal{T}\T P_{br}  \mathcal{T}^{-\mathsf{T}}, ~ \mathcal{B} \triangleq -\mathcal{P}_1 (C_s\T \otimes I) \hat{\bar{s}}, \\ &\mathcal{T} \triangleq \left(\begin{bmatrix} C_b C_{nb} \\ C_r C_{nb} \\ C_{ns}\end{bmatrix} \otimes I \right), ~~~~\mathcal{T}^{-\mathsf{T}} = \left(\begin{bmatrix}
    \frac{1}{2}C_b C_{nb} \\ 2 C_r C_{nb}\\ C_{ns}\end{bmatrix} \otimes I \right), \label{eq:Tdef}
\end{align}

\noindent where $\mathds{1} \triangleq [1~1~1]\T$ with $C_b$, $C_s$, $C_r$ and $C_{ns}$ being the connectivity matrices for bars, strings, center of the bars and point masses, respectively. The matrices $C_{nb} = [I~~0], M_{br} \triangleq blkdiag(M_b,M_r,M_{rs})$, $D_{br} \triangleq blkdiag(D_b,\mathbf{0},\mathbf{0})$, $K_{br} \triangleq blkdiag(K_b,\mathbf{0},\mathbf{0})$, and $P_{br} \triangleq blkdiag(P_b,\mathbf{I},\mathbf{I})$ are block diagonal matrices with $M_b \triangleq diag(J_1I, J_2I,...)$, $M_r \triangleq diag(m_1I, m_2I,...)$, $M_{rs} \triangleq diag(m_{s1}I, m_{s2}I,...)$, $D_b \triangleq diag(D_{b1},D_{b2},...)$, and $K_b \triangleq diag(K_{b1},K_{b2},...)$.
\begin{align}
    &D_{b1} \triangleq \frac{2J_1}{l_1^2}{\bar{b}_1} \dot{\bar{b}}_1\T, \\
    &K_{b1} \triangleq \left[\frac{J_1}{l_1^2}\dot{\bar{b}}_1\T \dot{\bar{b}}_1+\frac{1}{2l_1^2}\bar{b}_1\T (\bar{f}_2 - \bar{f}_1)\right]I+\frac{1}{2l_1^2}\bar{b}_1 (\bar{f}_2 - \bar{f}_1)\T, \\
    & P_{b1} \triangleq \frac{1}{2}\left(I- \frac{\bar{b}_1\bar{b}_1\T}{l_1^2}\right),
\end{align}

\noindent where $P_b \triangleq diag(P_{b1},P_{b2},...)$ with bar vector $\bar{b}_1$, string vector $\bar{s}_1$, bar velocity vector $\dot{\bar{b}}_1$, center of mass vector $\bar{r}_1$, center of mass velocity vector $\dot{\bar{r}}_1$, point mass vector $\bar{r}_{s1}$ and force $\bar{f}$, represents the vectors about which the dynamics is linearized.
\end{lemma}

\emph{\textbf{Proof:}}
The vector equations for rotational and translational dynamics of a bar are given as \cite{Goyal_Dynamics_2019}:
\begin{align}
J_1 \ddot{{b}}_1+\frac{J_1}{l_1^2}{b_1}\dot{{b_1}}\T \dot{{b_1}} &= \frac{1}{2}({f}_2 - {f}_1) - \frac{1}{2l_1^2}{b_1}{b_1}\T ({f}_2 - {f}_1),
\end{align}
\begin{align}
    m_1\ddot{r}_1 &= f_1 + f_2,
\end{align}

\noindent where $m_1$ and $J_1$ are the mass and inertia of bar about its mass center. The equation for a point mass with mass $m_{s1}$ (connecting string to string node) is given as \cite{Goyal_Dynamics_2019}:
\begin{align}
        m_{s1}\ddot{r}_{s1} &= f_{s1}.
\end{align}

Let us linearize these equations about some equilibrium bar vector $\bar{b}_1$, bar velocity vector $\dot{\bar{b}}_1$, center of mass vector $\bar{r}_1$, center of mass velocity vector $\dot{\bar{r}}_1$, point mass vector $\bar{r}_{s1}$ and force $\bar{f}$ such that:
\begin{multline}
\tilde{b}_1=b_1-\bar{b}_1,\tilde{r}_1=r_1-\bar{r}_1,\tilde{r}_{s1}=r_{s1}-\bar{r}_{s1}, \\ \dot{\tilde{b}}_1=\dot{b}_1-\dot{\bar{b}}_1,  \dot{\tilde{r}}_1=\dot{r}_1-\dot{\bar{r}}_1,\tilde{f}=f-\bar{f}.
\end{multline}

The linearized equations of motion for a bar and a point mass with $\tilde{b}_1,~\tilde{r}_1,~\tilde{r}_{s1}~\dot{\tilde{b}}_1,~ \dot{\tilde{r}}_1,~\dot{\tilde{r}}_{s1}$ being the linear variation can be written as:
\begin{multline}
J_1 \ddot{\tilde{b}}_1  + \frac{J_1}{l_1^2}\dot{\bar{b}}_1\T \dot{\bar{b}}_1\tilde{b}_1+\frac{2J_1}{l_1^2}{\bar{b}_1} \dot{\bar{b}}_1\T\dot{\tilde{b}}_1 = \frac{1}{2}(\tilde{f}_2 - \tilde{f}_1) - \frac{1}{2l_1^2}\bar{b}_1\T (\bar{f}_2 - \bar{f}_1)\tilde{b}_1 \\
- \frac{1}{2l_1^2}\bar{b}_1 (\bar{f}_2 - \bar{f}_1)\T\tilde{b}_1- \frac{1}{2l_1^2}\bar{b}_1\bar{b}_1\T (\tilde{f}_2 - \tilde{f}_1),
\end{multline}
\begin{align}
m_1\ddot{\tilde{r}}_1 &= \tilde{f}_1 + \tilde{f}_2, \\
m_{s1}\ddot{\tilde{r}}_{s1} &= \tilde{f}_{s1}.
\end{align}

Collecting the terms with $\ddot{\tilde{b}}_1,~\dot{\tilde{b}}_1,~ {\tilde{b}}_1,~ \tilde{f},~\ddot{\tilde{r}}_1$ and $\ddot{\tilde{r}}_{s1}$, we get:
\begin{multline}
\underbrace{\left[J_1I\right]}_{M_{b1}} \ddot{\tilde{b}}_1 +\underbrace{\left[\frac{2J_1}{l_1^2}{\bar{b}_1} \dot{\bar{b}}_1\T\right]}_{D_{b1}}\dot{\tilde{b}}_1  \\ +\underbrace{\left[ \left[\frac{J_1}{l_1^2}\dot{\bar{b}}_1\T \dot{\bar{b}}_1+\frac{1}{2l_1^2}\bar{b}_1\T (\bar{f}_2 - \bar{f}_1)\right]I+\frac{1}{2l_1^2}\bar{b}_1 (\bar{f}_2 - \bar{f}_1)\T\right]}_{K_{b1}}\tilde{b}_1 \\= \left[-\frac{1}{2}\left(I- \frac{\bar{b}_1\bar{b}_1\T}{l_1^2}\right)~~~\frac{1}{2}\underbrace{\left(I- \frac{\bar{b}_1\bar{b}_1\T}{l_1^2}\right)}_{P_{b1}}\right]\begin{bmatrix} {\tilde{f}}_1 \\ {\tilde{f}}_2 
\end{bmatrix},
\end{multline}
which again can be written using the assumed definitions as:
\begin{align}
    M_{b1} \ddot{\tilde{b}}_1 + D_{b1} \dot{\tilde{b}}_1 + K_{b1} \tilde{b}_1  = \frac{1}{2} P_{b1}\left[-I ~~~I\right] \begin{bmatrix} {\tilde{f}}_1 \\ {\tilde{f}}_2 
\end{bmatrix},
\end{align}
\begin{align}
\underbrace{\left[m_1I\right]}_{M_{r1}}\ddot{\tilde{r}}_1 = \left[I ~~~I\right]\begin{bmatrix} {\tilde{f}}_1 \\ {\tilde{f}}_2 
    \end{bmatrix}, \\
    \underbrace{[m_{s1}I]}_{M_{rs1}}\ddot{\tilde{r}}_{s1} = \tilde{f}_{s1}.
\end{align}

After linearizing the equation for a single bar and a point mass, let us now write the equations for all the bars together by stacking the bar vectors in one column and center of mass vectors in another column:
\begin{multline}
    \begin{bmatrix}
    M_{b1} & &  \\
    & M_{b2} &  \\
    & & \ddots  & \end{bmatrix}  \begin{bmatrix}
    \ddot{\tilde{b}}_1 \\ \ddot{\tilde{b}}_2  \\ \vdots
    \end{bmatrix}+ \begin{bmatrix}
    D_{b1} & &  \\
    & D_{b2} &  \\
    & & \ddots  & \end{bmatrix} \begin{bmatrix}
    \dot{\tilde{b}}_1 \\ \dot{\tilde{b}}_2  \\ \vdots
    \end{bmatrix}
    \\ +\begin{bmatrix}
    K_{b1} & &  \\
    & K_{b2} &  \\
    & & \ddots  & \end{bmatrix} \begin{bmatrix}
    {\tilde{b}}_1 \\ {\tilde{b}}_2  \\ \vdots
    \end{bmatrix} = \begin{bmatrix}
    P_{b1} & &  \\
    & P_{b2} &  \\
    & & \ddots  & \end{bmatrix} \left(\frac{C_b}{2} \otimes I \right) \tilde{f}_b,
\end{multline}
    
\begin{align}
    \begin{bmatrix}
    M_{r1} & &  \\
    & M_{r2} &  \\
    & & \ddots  & \end{bmatrix}  \begin{bmatrix}
    \ddot{\tilde{r}}_1 \\ \ddot{\tilde{r}}_2  \\ \vdots
    \end{bmatrix} &= \left(2 C_r \otimes I \right) \tilde{f}_b,
\end{align}
\begin{align}
    \begin{bmatrix}
    M_{rs1} & &  \\
    & M_{rs2} &  \\
    & & \ddots  & \end{bmatrix}  \begin{bmatrix}
    \ddot{\tilde{r}}_{s1} \\ \ddot{\tilde{r}}_{s2}  \\ \vdots
    \end{bmatrix} &=   \tilde{f}_s,
\end{align}
which again can be simply written using the definitions given in Lemma~\ref{L:linear_class1} as:
\begin{align}
        {M}_b \ddot{\tilde{b}} + {D}_b \dot{\tilde{b}} + {K}_b {\tilde{b}} &= {P}_b \left(\frac{1}{2}C_bC_{nb} \otimes I \right)\tilde{f}, \\
        {M}_r \ddot{\tilde{r}} &= \left(2 C_rC_{nb} \otimes I \right) \tilde{f}, \\
        {M}_{rs} \ddot{\tilde{r}}_s &= \left( C_{ns} \otimes I \right) \tilde{f}.
\end{align}

Let us stack column of bar vectors on top of the column of the center of mass vectors as: 
\begin{multline}
    \begin{bmatrix}
    {M}_b &  & \\
    & {M}_r & \\
    & & {M}_{rs} \end{bmatrix}  \begin{bmatrix}
    \ddot{\tilde{b}} \\ \ddot{\tilde{r}}  \\ \ddot{\tilde{r}}_s
    \end{bmatrix}+ \begin{bmatrix}
    {D}_b & & \\
    & \mathbf{0} & \\
    & & \mathbf{0}\end{bmatrix} \begin{bmatrix}
    \dot{\tilde{b}} \\ \dot{\tilde{r}} \\ \dot{\tilde{r}}_s
    \end{bmatrix}
    \\ +\begin{bmatrix}
    {K}_b & &\\
    & \mathbf{0} & \\
    & & \mathbf{0}\end{bmatrix} \begin{bmatrix}
    {\tilde{b}} \\ {\tilde{r}}  \\ {\tilde{r}}_s 
    \end{bmatrix}  = \begin{bmatrix}
    {P_b}  & &\\
      & I & \\
    & &  I\end{bmatrix} \left(\begin{bmatrix}
    \frac{1}{2}C_b C_{nb} \\ 2 C_r C_{nb}\\ C_{ns}\end{bmatrix} \otimes I \right)     \tilde{f}.
    \label{eq:Linear_M_bM_r}
\end{multline}

Using $ \begin{bmatrix} \tilde{b} \\ \tilde{r} \\ \tilde{r}_s \end{bmatrix} = \mathcal{T} \tilde{n}$ with $\mathcal{T}$ given in equation (\ref{eq:Tdef}) and 
% \begin{align}
%     \mathcal{T} \triangleq \left(\begin{bmatrix} C_b C_{nb} \\ C_r C_{nb} \\ C_{ns}\end{bmatrix} \otimes I \right),
% \end{align}
% we see that 
% \begin{align}
%     \mathcal{T}^{-\mathsf{T}} = \left(\begin{bmatrix}
%     \frac{1}{2}C_b C_{nb} \\ 2 C_r C_{nb}\\ C_{ns}\end{bmatrix} \otimes I \right).
% \end{align}
substituting it in equation (\ref{eq:Linear_M_bM_r}), we get:
\begin{align}
    M_{br}  \mathcal{T}\ddot{\tilde{n}}  + D_{br} \mathcal{T}\dot{\tilde{n}}
    +K_{br}\mathcal{T}{\tilde{n}} = P_{br} \mathcal{T}^{-\mathsf{T}} \tilde{f}.
\end{align}

Multiplying from the left-hand side by $\mathcal{T}\T$, we get:
\begin{align}
   \mathcal{T}\T M_{br}  \mathcal{T}\ddot{\tilde{n}}  + \mathcal{T}\T D_{br} \mathcal{T}\dot{\tilde{n}}
    +\mathcal{T}\T K_{br}\mathcal{T}{\tilde{n}} = \mathcal{T}\T P_{br} \mathcal{T}^{-\mathsf{T}} \tilde{f},
\end{align}
which after defining new variables can be written as:
\begin{align}
    \mathcal{M} \ddot{\tilde{n}} + \mathcal{D} \dot{\tilde{n}} + \mathcal{K} \tilde{n} = \mathcal{P}\tilde{f}.
        \label{eq:Linear_n_f}
\end{align}

\noindent \textbf{\textit{Force density $\gamma$ as control variable:}}
The above formulated linearized dynamics is only for bars in the presence of some external force $f$. In order to include the forces due to tension in the strings, we can divide the force $f$ into two parts: one as external force $w$ and the other as internal forces due to strings tension as $t$ (actually formulated as $\gamma$). 
From the nonlinear dynamics model development given in \cite{Goyal_Dynamics_2019}, we write the following in the vector form as:
% \begin{gather}
%     F = W - S\hat{\gamma}C_s ,
% \end{gather}
% which can be written in vector form as:
\begin{align}
        f = w - (C_s\T \otimes I) \widehat{(\gamma\otimes \mathds{1})} s,
\end{align}
where $\mathds{1} = [1~1~1]\T$. Linearizing the above equation about $\tilde{\gamma} = \gamma - \bar{\gamma},~ \tilde{w} = w - \bar{w},~ \tilde{s} = s - \bar{s}$, and $\dot{\tilde{s}}=\dot{s}-\dot{\bar{s}}$, we get:
\begin{align}
    \tilde{f} = \tilde{w} - \underbrace{(C_s\T \otimes I) \widehat{(\bar{\gamma} \otimes \mathds{1})}}_{K_s} \tilde{s} - \underbrace{(C_s\T \otimes I) \hat{\bar{s}}}_{K_\gamma}\tilde{\gamma},
\end{align}
\begin{align}
    \tilde{f} = \tilde{w} - {K_s} \tilde{s} - {K_\gamma}\tilde{\gamma}.
    \label{eq:Linear_f_wGamma}
\end{align}

Substituting the above equation in equation (\ref{eq:Linear_n_f}), we get:
\begin{align}
    \mathcal{M} \ddot{\tilde{n}} + \mathcal{D} \dot{\tilde{n}} + \mathcal{K} \tilde{n} = \mathcal{P}\tilde{w} - \mathcal{P}{K_s} \tilde{s} -\mathcal{P} {K_\gamma}\tilde{\gamma}.
\end{align}

Using $\tilde{s} = (C_s \otimes I)\tilde{n}$, the above equation can easily be converted to $\tilde{n}$ coordinates as:
\begin{align}
    \mathcal{M} \ddot{\tilde{n}} + \mathcal{D} \dot{\tilde{n}} + (\mathcal{K}+ \mathcal{P}{K_s} (C_s \otimes I)) \tilde{n} = \mathcal{P}\tilde{w} -\mathcal{P} {K_\gamma}\tilde{\gamma},
    \label{eq:Linear_n_gamma}
\end{align}
\begin{align}
    \mathcal{M}_1 \ddot{\tilde{n}} + \mathcal{D}_1 \dot{\tilde{n}} + \mathcal{K}_1 \tilde{n} = \mathcal{P}_1 \tilde{w}+\mathcal{B}_1 \tilde{\gamma}.
\end{align}

% \begin{align}
%     \mathbf{M}_1\ddot{\tilde{n}} + \mathbf{D}_1\dot{\tilde{n}} + \mathbf{K}_1{\tilde{n}} = \mathbf{B}_1 \tilde{w} + \mathbf{B}_2 \tilde{\gamma}.
% \end{align}
% where 
% \begin{align}
%     \mathbf{M}_1 = M_{br}\left(\begin{bmatrix} C_b \\ C_r \\ C_{ns}\end{bmatrix} \otimes I \right), \hspace{0.5 pc} \mathbf{D}_1 = D_{br}\left(\begin{bmatrix} C_b \\ C_r \\ C_{ns}\end{bmatrix} \otimes I \right), \hspace{0.5 pc} &\mathbf{K}_1 = K_{br}\left(\begin{bmatrix} C_b \\ C_r \\ C_{ns}\end{bmatrix} \otimes I  \right) + P_{br} K_s (C_s \otimes I), \\
%     \mathbf{B}_1 = P_{br}, \hspace{1pc}& \mathbf{B}_2 = - P_{br} K_\gamma
% \end{align}

% The above equation can be written in state space form as:
% \begin{align}
%     \begin{bmatrix}
%     \dot{\tilde{n}} \\ \ddot{\tilde{n}} 
%     \end{bmatrix}  = \begin{bmatrix}
%     0 & I \\
%     -\textbf{M}_1^{-1}\textbf{K}_1 &-\textbf{M}_1^{-1}\textbf{D}_1
%     \end{bmatrix} \begin{bmatrix}
%     \tilde{n} \\ \dot{\tilde{n}}
%     \end{bmatrix} + \begin{bmatrix}
%     0 & 0 \\ \textbf{M}_1^{-1} \mathbf{B}_1 &\textbf{M}_1^{-1}\textbf{B}_2
%     \end{bmatrix}
%     \begin{bmatrix}
%     \tilde{w} \\ \tilde{\gamma}
%     \end{bmatrix} 
% \end{align}
\hfill \qedsymbol

Notice that the above equation is only applicable for closed-loop control where we assume the force density $\tilde{\gamma}$ to be the control input. In physical problems, the control input may be the rest length of the strings. \\

\subsubsection{String rest length $\rho$ as the control variable - Linearizing force density $\gamma$}
The above two subsections are formulated for closed-loop dynamics where $\gamma$ is defined as input control variable. In this subsection, we convert $\gamma$ to rest length to run open-loop dynamics simulations. Assuming that strings follow Hooke's law and viscous friction damping model, the tension in a string is written as:
\begin{align}
&\|t_i\|=k_i(\|s_i\|-\rho_i)+c_i\frac{s_i\T\dot{s_i}}{\|s_i\|},\\
&\gamma_i=\frac{\|t_i\|}{\|s_i\|}=k_i\left(1-\frac{\rho_i}{\|s_i\|}\right)+c_i\frac{s_i\T\dot{s_i}}{\|s_i\|^2},
\end{align}
where $\rho_i$ is rest length of the string, $k_i$ is the extensional stiffness, $c_i$ is the damping coefficient. The force density $\gamma_i$ can again be written as:
\begin{align}
    \gamma_i=k_i\left(1-\frac{\rho_i}{(s_i\T s_i)^{1/2}}\right)+c_i\frac{s_i\T\dot{s_i}}{s_i\T s_i}.
\end{align}

Linearizing the force density $\gamma$ about equilibrium values ($\tilde{\gamma} = \gamma - \bar{\gamma},~\tilde{\rho} = \rho - \bar{\rho},~ \tilde{s} = s - \bar{s}$, and  $\dot{\tilde{s}} = \dot{s} - \dot{\bar{s}}$), we get:
\begin{multline}
    \tilde{\gamma}_i = \underbrace{\left[\frac{k_i \bar{\rho_i} \bar{s}_i\T}{(\bar{s_i}\T \bar{s_i})^{3/2}}+\frac{c_i \dot{\bar{s}}_i\T}{\bar{s_i}\T \bar{s_i}}-\frac{2 c_i \dot{\bar{s}}_i\T \bar{s}_i \bar{s}_i\T }{(\bar{s_i}\T \bar{s_i})^{2}}\right]}_{\zeta}\tilde{s}_i \\ +\underbrace{\left[\frac{c_i {\bar{s}}_i\T}{\bar{s_i}\T \bar{s_i}}\right]}_{\kappa}\dot{\tilde{s}}_i - \underbrace{\frac{k_i}{(\bar{s_i}\T \bar{s_i})^{1/2}}}_{\iota} \tilde{\rho_i}.
\end{multline}

Stacking the force densities for all the strings from the above equation will give:
\begin{align}
    \tilde{\gamma} &= \begin{bmatrix}
    \zeta_1 & & \\
    & \zeta_2 & \\
    & & \ddots
    \end{bmatrix} \tilde{s} + \begin{bmatrix}
    \kappa_1 & & \\
    & \kappa_2 & \\
    & & \ddots
    \end{bmatrix} \dot{\tilde{s}} - \begin{bmatrix}
    \iota_1 & & \\
    & \iota_2 & \\
    & & \ddots
    \end{bmatrix} \tilde{\rho}, \\
    \tilde{\gamma} &= K_{ks}\tilde{s} + K_{cs} \dot{\tilde{s}} - K_{ps} \tilde{\rho}.
\end{align}

The above equation can be substituted to final linearized equation (\ref{eq:Linear_n_gamma}) as:
\begin{multline}
    \mathcal{M} \ddot{\tilde{n}} + \mathcal{D} \dot{\tilde{n}} + (\mathcal{K}+ \mathcal{P}{K_s} (C_s \otimes I)) \tilde{n} = \mathcal{P}\tilde{w} \\ -\mathcal{P} {K_\gamma}(K_{ks}\tilde{s} + K_{cs} \dot{\tilde{s}} - K_{ps} \tilde{\rho} ).
\end{multline}

% \begin{align}
%     M_{br}  \begin{bmatrix}
%     \ddot{\tilde{b}} \\ \ddot{\tilde{r}} \\  \ddot{\tilde{r}}_s
%     \end{bmatrix} + D_{br} \begin{bmatrix}
%     \dot{\tilde{b}} \\ \dot{\tilde{r}} \\  \dot{\tilde{r}}_s 
%     \end{bmatrix}
%     +K_{br} \begin{bmatrix}
%     {\tilde{b}} \\ {\tilde{r}} \\  {\tilde{r}}_s
%     \end{bmatrix} = P_{br}\tilde{w} - P_{br} K_{s} \tilde{s} - P_{br} K_\gamma  K_{ks}\tilde{s} - P_{br} K_\gamma  K_{cs}\dot{\tilde{s}}
% \end{align}

Again using $\tilde{s} = (C_s \otimes I)\tilde{n}$, the above equation can easily be converted to $\tilde{n}$ coordinates as:

\begin{multline}
    \mathcal{M} \ddot{\tilde{n}} + (\mathcal{D}+\mathcal{P} {K_\gamma}K_{cs}(C_s \otimes I)) \dot{\tilde{n}} + (\mathcal{K}+ \mathcal{P}{K_s} (C_s \otimes I) \\ +\mathcal{P} {K_\gamma}K_{ks}(C_s \otimes I)) \tilde{n} = \mathcal{P}\tilde{w}+\mathcal{P} {K_\gamma}K_{ps}\tilde{\rho}.
\end{multline}
This equation represents the linearized dynamics model with rest length $\tilde{\rho}$ as the control input around the equilibrium point. 
The following equation can be obtained for the open-loop linearized dynamics by substituting for the $\tilde{\rho} = 0$:
\begin{multline}
    \mathcal{M} \ddot{\tilde{n}} + (\mathcal{D}+\mathcal{P} {K_\gamma}K_{cs}(C_s \otimes I)) \dot{\tilde{n}} \\  + (\mathcal{K}+ \mathcal{P}{K_s} (C_s \otimes I)+\mathcal{P} {K_\gamma}K_{ks}(C_s \otimes I)) \tilde{n} = \mathcal{P}\tilde{w}.
\end{multline}
The above equation will result in the motion only affected by perturbation in the external force and will assume the rest length of the strings to be constant.

% \begin{align}
%     \mathbf{M}\ddot{\tilde{n}} + \mathbf{D}\dot{\tilde{n}} + \mathbf{K}{\tilde{n}} = \mathbf{B} \tilde{w} 
% \end{align}
% where 
% \begin{align}
%     \mathbf{M} = M_{br}\left(\begin{bmatrix} C_b \\ C_r \\ C_{ns}\end{bmatrix} \otimes I \right), \hspace{0.5 pc} \mathbf{D} = D_{br}\left(\begin{bmatrix} C_b \\ C_r \\ C_{ns}\end{bmatrix} \otimes I \right) + P_{br} K_\gamma K_{cs} (C_s \otimes I),\\ \mathbf{K} = K_{br}\left(\begin{bmatrix} C_b \\ C_r \\ C_{ns}\end{bmatrix} \otimes I  \right) + P_{br} K_s (C_s \otimes I)+ P_{br} K_\gamma K_{ks} (C_s \otimes I), \hspace{0.5 pc} 
%     \mathbf{B} = P_{br}
% \end{align}

% and the final state space form can be written as:
% \begin{align}
%     \begin{bmatrix}
%     \dot{\tilde{n}} \\ \ddot{\tilde{n}} 
%     \end{bmatrix}  = \begin{bmatrix}
%     0 & I \\
%     -\textbf{M}^{-1}\textbf{K} &-\textbf{M}^{-1}\textbf{D}
%     \end{bmatrix} \begin{bmatrix}
%     \tilde{n} \\ \dot{\tilde{n}}
%     \end{bmatrix} + \begin{bmatrix}
%     0  \\ \textbf{M}^{-1} \mathbf{B}
%     \end{bmatrix}
%     \tilde{w} 
% \end{align}

\subsection{Reduced-order Linearized Dynamics for Class-k Systems}
The linearization of class-k dynamics considers linear constraints in addition to the above-mentioned linearized dynamics developed for the class-1 systems. Please refer to \cite{Goyal_Dynamics_2019} for the complete reduced-order nonlinear dynamics model of the class-k tensegrity structure.

\begin{lemma}
The reduced-order linearized dynamics of any class-k tensegrity system can be written as:

\begin{align}
    \mathcal{M}_k \ddot{\tilde{\eta}}_2 + \mathcal{D}_k \dot{\tilde{\eta}}_2 + \mathcal{K}_k \tilde{\eta}_2 = \mathcal{P}_k\tilde{w}+\mathcal{B}_k
    \tilde{\gamma},
    \label{eq:class-k_linear_dyn}
\end{align}
where $\mathcal{M}_k \triangleq V_2\T \mathcal{M}_1 V_2,  \mathcal{D}_k \triangleq V_2\T\mathcal{D}_1 V_2, \mathcal{K}_k \triangleq V_2\T \mathcal{K}_1 V_2, \mathcal{P}_k \triangleq V_2\T\mathcal{P}_1, \mathcal{B}_k \triangleq V_2\T\mathcal{B}_1$, with some linear constraints of the form for class-k structure as:

\begin{align}
    % A \bar{n} = d, \hspace{0.5pc} 
    \nonumber A \tilde{n} =0, ~U \begin{bmatrix}
    \Sigma_1 & 0
    \end{bmatrix}\begin{bmatrix}
    V_1\T \\ V_2\T 
    \end{bmatrix}  \tilde{n} = U \begin{bmatrix}
    \Sigma_1 & 0 
    \end{bmatrix}     \begin{bmatrix} \tilde{\eta}_1 \\ \tilde{\eta}_2 \end{bmatrix}=0,
    % \label{eq:constraint_linear}
\end{align}
implying $\tilde{\eta}_1=\dot{\tilde{\eta}}_1=\ddot{\tilde{\eta}}_1=0.$
\end{lemma}

\emph{\textbf{Proof:}}

Let us start with writing the linearized dynamics equation for bars (equation (\ref{eq:Linear_n_f})) here again as:
\begin{align}
    \mathcal{M} \ddot{\tilde{n}} + \mathcal{D} \dot{\tilde{n}} + \mathcal{K} \tilde{n} = \mathcal{P}\tilde{f}.
\end{align}

For class-k structure, we have linear constraints of the form (refer to \cite{Goyal_Dynamics_2019} for more details):
\begin{align}
    A \bar{n} = d, \hspace{1pc} A \tilde{n} = 0,
        \label{eq:constraint_linear}
\end{align}
with $A \in\mathbb{R}^{N_c\times 3n}$ and $N_c$ representing the number of constraints.
In the presence of these constraints, the new dynamics can be written as:
\begin{align}
    \mathcal{M} \ddot{\tilde{n}} + \mathcal{D} \dot{\tilde{n}} + \mathcal{K} \tilde{n} = \mathcal{P}\tilde{f} + A\T \Omega,
    \label{eq:classk_dyn_linear}
\end{align}
where $\Omega$ is the Lagrange multiplier and $A\T \Omega$ represents the constraint force. Adding the linear constraints into the dynamics will restrict the motion in certain dimensions, thus reducing the order of the dynamics to a span a smaller space. To this end, we use the singular value decomposition (SVD) of matrix $A$ as:

\begin{align}
A=U\Sigma V\T=U \begin{bmatrix}
    \Sigma_1 & 0
    \end{bmatrix}\begin{bmatrix}
    V_1\T \\ V_2\T 
    \end{bmatrix} ,
\label{PSVD_linear}
\end{align}
where $U\in\mathbb{R}^{N_c\times N_c}$ and $V\in\mathbb{R}^{3n\times 3n}$ are both unitary matrices, $V_1\in\mathbb{R}^{3n\times N_c}$ and $V_2\in\mathbb{R}^{3n\times(3n-N_c)}$ are submatrices of $V$, and $\Sigma_1\in\mathbb{R}^{N_c\times N_c}$ is a diagonal matrix  of positive singular values. By defining
\begin{align}
\eta=\begin{bmatrix} \eta_1 \\ \eta_2 \end{bmatrix} \triangleq V\T \tilde{n} = \begin{bmatrix} V_1\T \tilde{n} \\ V_2\T \tilde{n} \end{bmatrix},
\end{align}
the constraint equation (\ref{eq:constraint_linear}) can be modified as:
\begin{align}
A \tilde{n} =U\Sigma V\T \tilde{n} = U \begin{bmatrix}
    \Sigma_1 & 0 
    \end{bmatrix}     \begin{bmatrix} \eta_1 \\ \eta_2 \end{bmatrix}=0,
\end{align}
which implies:
\begin{align}
\eta_1=0, \;\;\dot{\eta}_1=0,\;\;\ddot{\eta}_1=0.\label{eta_1_linear}
\end{align}
Notice that the $\eta_1$-space represents the no-motion space in transformed coordinates and the $\eta_2$-space will evolve according to the constrained dynamics in new coordinate system. Using Equations (\ref{PSVD_linear}-\ref{eta_1_linear}), the dynamics equation (\ref{eq:classk_dyn_linear}) can be rewritten as:

\begin{align}
\mathcal{M}V_2 \ddot{\eta_2}+\mathcal{D}V_2 \dot{\eta_2}+\mathcal{K} V_2 \eta_2 = \mathcal{P}\tilde{f} + V_1 \Sigma_1 U\T \Omega. 
\end{align}

Pre-multiplying the above equation by a non-singular matrix $[V_1~~V_2 ]\T$ will  yield two parts, where second part gives the second order differential equation for the reduced dynamics:
\begin{align}
\nonumber V_2\T  \mathcal{M}V_2 \ddot{\eta_2}+V_2\T  \mathcal{D}V_2 \dot{\eta_2}+V_2\T \mathcal{K} V_2 \eta_2  &= V_2\T \mathcal{P}\tilde{f} + V_2\T
V_1 \Sigma_1 U\T \Omega ,
\end{align}
\begin{align}
\Rightarrow \mathcal{M}_2 \ddot{\eta}_2+\mathcal{D}_2 \dot{\eta}_2+\mathcal{K}_2\eta_2&= \mathcal{P}_2 \tilde{f}.
\label{eq:RedDynU1_linear}
\end{align}
with $\mathcal{M}_2 = V_2\T \mathcal{M}V_2,\mathcal{D}_2 = V_2\T \mathcal{D}V_2$, $\mathcal{K}_2 = V_2\T  \mathcal{K}V_2$, and $\mathcal{P}_2 = V_2\T  \mathcal{P}$.\\

Let us substitute $\tilde{f}$ in terms of $\tilde{w}$ and $\tilde{\gamma}$ from equation (\ref{eq:Linear_f_wGamma}) to incorporate the string forces:
\begin{align}
    \mathcal{M}_2 \ddot{\eta}_2 +  \mathcal{D}_2 \dot{\eta}_2 +  \mathcal{K}_2 \eta_2 =  \mathcal{P}_2\tilde{w} -  \mathcal{P}_2 {K_s} \tilde{s} - \mathcal{P}_2 {K_\gamma}\tilde{\gamma},
\end{align}

Using $\tilde{s} = (C_s \otimes I)\tilde{n} = (C_s \otimes I)V_2 \eta_2$, the above equation can easily be converted to $\tilde{n}$ coordinates as:
\begin{align}
    \mathcal{M}_2 \ddot{\eta}_2 +  \mathcal{D}_2 \dot{\eta}_2 + (\mathcal{K}_2+ \mathcal{P}_2{K_s} (C_s \otimes I)V_2) \eta_2 =  \mathcal{P}_2\tilde{w} - \mathcal{P}_2 {K_\gamma}\tilde{\gamma},
\end{align}
\begin{align}
    \mathcal{M}_k \ddot{\eta}_2 +  \mathcal{D}_k \dot{\eta}_2 + \mathcal{K}_k \eta_2 =  \mathcal{P}_k\tilde{w} + \mathcal{B}_k \tilde{\gamma}.
\end{align}
This completes the proof. \hfill \qedsymbol

Now using $\tilde{\gamma} = K_{ks}\tilde{s} + K_{cs} \dot{\tilde{s}}-K_{ps}\tilde{\rho}$, and again using $\tilde{s} = (C_s \otimes I)V_2 \eta_2$, the above equation can easily be converted to $\tilde{n}$ coordinates as:
\begin{multline}
    \mathcal{M}_2 \ddot{\eta}_2 + (\mathcal{D}_2 +\mathcal{P}_2 {K_\gamma}K_{cs}(C_s \otimes I)V_2) \dot{\eta}_2 + (\mathcal{K}_2+ \mathcal{P}_2 {K_s} (C_s \otimes I)V_2 \\ +\mathcal{P}_2 {K_\gamma}K_{ks}(C_s \otimes I)V_2) \eta_2  =  \mathcal{P}_2 \tilde{w}+ \mathcal{P}_2 {K_\gamma}K_{ps}\tilde{\rho}.
\end{multline}

The following equation can be obtained for the open-loop linearized dynamics by substituting for the $\tilde{\rho} = 0$:
\begin{multline}
    \mathcal{M}_2 \ddot{\eta}_2 + (\mathcal{D}_2 +\mathcal{P}_2 {K_\gamma}K_{cs}(C_s \otimes I)V_2) \dot{\eta}_2 + (\mathcal{K}_2+ \mathcal{P}_2 {K_s} (C_s \otimes I)V_2 \\ +\mathcal{P}_2 {K_\gamma}K_{ks}(C_s \otimes I)V_2) \eta_2  =  \mathcal{P}_2 \tilde{w}.
\end{multline} 

This section developed the linearized model about an equilibrium point for the class-1 and class-k tensegrity structures. The bar length constraint present in the actual nonlinear dynamics model is lost during the linearization process, resulting in a maximal-coordinates representation with some uncontrollable modes.

\section{Minimal-coordinates representation of the linearized system}
The purpose of this section is to provide a novel formulation to remove the uncontrollable modes which correspond to the bar length change. As the nonlinear dynamical model assumes bars in the system to be rigid, it is imperative to constrain the bar length change by the removal of such modes for a physically realizable system. This reduction in the order of the system is shown to generate a minimal-coordinates representation of the system for both class-1 and class-k tensegrity systems.

\subsection{Class-1 Tensegrity System Dynamics}\label{ss:class1_min}

To generate the minimal order linearized dynamics model for both class-1 and class-k tensegrity system, we start with the linearized model for the class-1 system given in equation (\ref{eq:Linear_n_f}):

\begin{align}
    \mathcal{M} \ddot{\tilde{n}} + \mathcal{D} \dot{\tilde{n}} + \mathcal{K} \tilde{n} = \mathcal{P} \tilde{f},
\end{align}
and look for the modes in which the length of the bar is not changing. Let us start by writing the coordinate transformation from bar coordinates to nodes coordinates as: 
\begin{align}
    \tilde{n} = \left(\begin{bmatrix}
    \frac{1}{2}C_{nb}\T C_b\T  & 2 C_{nb}\T C_r\T & C_{ns}\T \end{bmatrix} \otimes I \right)\begin{bmatrix} \tilde{b} \\ \tilde{r} \\ \tilde{r}_s \end{bmatrix}
\end{align}

Now, the mode corresponding to the length change for all the bars can be calculated using $\tilde{r} = \tilde{r}_s = 0$ and $\tilde{b} = \delta \bar{b}$. This can be realized in nodal coordinates as:

\begin{align}
    \delta \bar{n} = \left(\begin{bmatrix}
    \frac{1}{2}C_{nb}\T C_b\T  & 2 C_{nb}\T C_r\T & C_{ns}\T \end{bmatrix} \otimes I \right)\begin{bmatrix} \delta \bar{b} \\ 0 \\ 0\end{bmatrix}.
\end{align}
Note that the mode $\delta \bar{n}$ represents the motion in the linearized mode where the length of all the bars are changing by a fraction of $\delta$. The modes corresponding to the length change of the $i$\textsuperscript{th} bar can be found by extracting the nodes related to that bar to create a $\delta \bar{n}_{1_i} \in \mathcal{R}^{6 \times 1}$ vector. Now, we find 5 modes which are perpendicular to this mode corresponding to the $i$\textsuperscript{th} bar by finding the right null space for the $\delta \bar{n}_{1_i}\T$ dimensional vector as $\delta \bar{n}_{2_i} =~ ^{\perp} \delta \bar{n}_{1_i}\T \in \mathcal{R}^{6 \times 5}$.
Now, the nodes can be arranged in their respective order to generate $\Phi_{1_i}\in \mathcal{R}^{6\beta \times 1}$ from $\bar{n}_{1_i}$ and $\Phi_{2_i}\in \mathcal{R}^{6\beta \times 5}$ from $\bar{n}_{2_i}$. 
The similar procedure can be done for all the bars to create $\Phi_1 = \begin{bmatrix}
\cdots \Phi_{1_i} \cdots \end{bmatrix} \in \mathcal{R}^{6\beta \times \beta} $ and $\Phi_2 = \begin{bmatrix} \cdots \Phi_{2_i} \cdots \end{bmatrix} \in \mathcal{R}^{6\beta \times 5\beta}$. Finally, the coordinate transformation matrix $\Phi = [\Phi_1 ~ \Phi_2] \in \mathcal{R}^{6\beta \times 6\beta}$ can be formulated which spans the entire $6\beta$ dimensional space. Also, notice that each column of the matrix $\Phi$ can be scaled to have unit length which will result in the matrix $\Phi$ to be orthonormal, i.e., $\Phi^{-1} = \Phi\T$.

Now, let us transform the coordinates from $\tilde{n}$ space to a new space $\phi$ space with $\tilde{n} = \Phi \phi$ and substituting it into equation (\ref{eq:Linear_n_f}) to obtain:
\begin{align}
    \mathcal{M} \Phi \ddot{\phi} + \mathcal{D} \Phi \dot{\phi} + \mathcal{K} \Phi \phi = \mathcal{P} \tilde{f},
\end{align}
which after multiplying from the left hand side by $\Phi\T$ can be written as:
\begin{align}
    \Phi\T\mathcal{M} \Phi \ddot{\phi} + \Phi\T\mathcal{D} \Phi \dot{\phi} + \Phi\T\mathcal{K} \Phi \phi = \Phi\T\mathcal{P} \tilde{f}.
\end{align}

Note that the feasible modes for the dynamics system with rigid bars are corresponding to $\Phi_2$ and thus, the physically feasible reduced-order system can be written as:

\begin{align}
    \Phi_2\T\mathcal{M} \Phi_2 \ddot{\phi} + \Phi_2\T \mathcal{D} \Phi_2 \dot{\phi} + \Phi_2\T \mathcal{K} \Phi_2 \phi = \Phi_2\T\mathcal{P}\tilde{f},
\end{align}
which after substituting for $\tilde{f}$ can be written as:
\begin{multline}
    \Phi_2\T\mathcal{M} \Phi_2 \ddot{\phi} + \Phi_2\T \mathcal{D} \Phi_2 \dot{\phi} + \Phi_2\T (\mathcal{K}+\mathcal{P} {K_s} (C_s \otimes I)) \Phi_2  {\eta}_{2_\phi} \\ =  \Phi_2\T\mathcal{P}\tilde{w} - \Phi_2\T\mathcal{P}{K_\gamma}\tilde{\gamma}.
\end{multline}

\noindent The matrix $\Phi_2\T \mathcal{K}_1 \Phi_2$ will be positive definite with feasible modes, corresponding to 3 translational and 2 rotational motion for each bar.

\subsection{Class-k Tensegrity System Dynamics}
The linear constraint for the class-k system should be added corresponding to the transformed coordinates $\phi$ to only keep the feasible modes. It was mentioned earlier that the linear constraints in the vector form could be written as:
\begin{align}
  A\tilde{n} = 0, 
%   \hspace{1pc} A = P\T \otimes I,
\end{align}
which can now be written in transformed coordinates as:
\begin{align}
  A_2~\phi = 0, \hspace{1pc} A_2 \triangleq A \Phi_2,
%   ~~ A = P\T \otimes I.
\end{align}

Now, following the same procedure as described earlier, we can write the dynamics in even more reduced-order as:

\begin{align}
    \Phi_2\T\mathcal{M} \Phi_2 \ddot{\phi} + \Phi_2\T \mathcal{D} \Phi_2 \dot{\phi} + \Phi_2\T \mathcal{K} \Phi_2 \phi = \Phi_2\T\mathcal{P}\tilde{f}+ A_2\T \Omega.
\end{align}
where $A_2$ can be decomposed as:
\begin{align}
A_2=U_\phi \Sigma_\phi V_\phi\T=U_\phi \begin{bmatrix}
    \Sigma_{1_\phi} & 0
    \end{bmatrix}\begin{bmatrix}
    V_{1_\phi}\T \\ V_{2_\phi}\T 
    \end{bmatrix} ,
\end{align}
and by defining:
\begin{align}
\eta_\phi = \begin{bmatrix} \eta_{1_\phi} \\ \eta_{2_\phi} \end{bmatrix} \triangleq V_\phi\T \phi = \begin{bmatrix} V_{1_\phi}\T \phi \\ V_{2_\phi}\T \phi \end{bmatrix} ,
\end{align}
the constraints can be modified as:
\begin{align}
A_2 \phi  = U_\phi \begin{bmatrix}
    \Sigma_{1_\phi} & 0 
    \end{bmatrix}     \begin{bmatrix} \eta_{1_\phi} \\ \eta_{2_\phi} \end{bmatrix}=0,
\end{align}
implying:
\begin{align}
\eta_{1_\phi}=0, \;\;\dot{\eta}_{1_\phi}=0,\;\;\ddot{\eta}_{1_\phi}=0.
\end{align}

Now, the dynamics with the constrained system and no bar length change will evolve as:
\begin{multline}
    \Phi_2\T\mathcal{M} \Phi_2 V_{2_\phi} \ddot{\eta}_{2_\phi} + \Phi_2\T \mathcal{D} \Phi_2 V_{2_\phi} \dot{\eta}_{2_\phi}+ \Phi_2\T \mathcal{K} \Phi_2 V_{2_\phi} {\eta}_{2_\phi} \\ = \Phi_2\T\mathcal{P}\tilde{f}+ V_{1_\phi} \Sigma_{1_\phi} U_\phi\T \Omega,
\end{multline}
and pre-multiplying the above equation by a non-singular matrix $[V_{1_\phi}~~V_{2_\phi} ]\T$ will  yield two parts, where second part gives the second order differential equation for the reduced dynamics:

\begin{multline}
   V_{2_\phi}\T \Phi_2\T\mathcal{M} \Phi_2 V_{2_\phi} \ddot{\eta}_{2_\phi} + V_{2_\phi}\T \Phi_2\T \mathcal{D} \Phi_2 V_{2_\phi} \dot{\eta}_{2_\phi}+ V_{2_\phi}\T \Phi_2\T \mathcal{K} \Phi_2 V_{2_\phi} {\eta}_{2_\phi} \\ = V_{2_\phi}\T \Phi_2\T\mathcal{P}\tilde{f}+ V_{2_\phi}\T V_{1_\phi} \Sigma_{1_\phi} U_\phi\T \Omega,
\end{multline}
which can finally be written as:
\begin{multline}
   V_{2_\phi}\T \Phi_2\T\mathcal{M} \Phi_2 V_{2_\phi} \ddot{\eta}_{2_\phi} + V_{2_\phi}\T \Phi_2\T \mathcal{D} \Phi_2 V_{2_\phi} \dot{\eta}_{2_\phi}+ V_{2_\phi}\T \Phi_2\T \mathcal{K} \Phi_2 V_{2_\phi} {\eta}_{2_\phi} \\ = V_{2_\phi}\T \Phi_2\T\mathcal{P}\tilde{f},
\end{multline}
and after substituting for $\tilde{f}$ in terms of $\tilde{w}$ and $\tilde{\gamma}$ from equation (\ref{eq:Linear_f_wGamma}) to incorporate the string forces:
\begin{multline}
   V_{2_\phi}\T \Phi_2\T\mathcal{M} \Phi_2 V_{2_\phi} \ddot{\eta}_{2_\phi} + V_{2_\phi}\T \Phi_2\T \mathcal{D} \Phi_2 V_{2_\phi} \dot{\eta}_{2_\phi} \\ + V_{2_\phi}\T \Phi_2\T (\mathcal{K}+\mathcal{P} {K_s} (C_s \otimes I)) \Phi_2 V_{2_\phi} {\eta}_{2_\phi} \\ = V_{2_\phi}\T \Phi_2\T\mathcal{P}\tilde{w} - V_{2_\phi}\T \Phi_2\T\mathcal{P}{K_\gamma}\tilde{\gamma}.
\end{multline}
This is the final linearized minimal-order dynamics equation for the class-k tensegrity system with force density in the strings $\tilde{\gamma}$ as the control variable.

\section{Closed-Loop response with different performance objectives using the linearized model}
In this section, the best performance bounds using different objectives are calculated for the closed-loop response of the linearized tensegrity system dynamics. The different performance criteria namely; covariance bound, bound on generalized $\mathcal{H}_2$ problem, bound on $\Gamma_{ie}$ and bound on $\mathcal{H}_\infty$ are considered to design a dynamic controller for the linearized system. The variation in the best performance bound with the controller in the loop is observed with varying the structure parameter (initial prestress value) of the tensegrity system. The problem of designing a dynamic controller with fixed structure parameters for the above-mentioned performance objectives is known to be a convex problem with a solution written in the form of Linear matrix inequalities (LMIs) \cite{Scherer_LMI_1997, Skelton_LMI_1998}. For the case of fixed structure parameters, these LMIs provide controller matrices and the best performance bound on the error. The purpose here is to understand the dependence of structure parameters on performance objectives for given values of actuator and sensor precision and then, the next section (Section V) provides the framework to simultaneously optimize structure, controller, and actuator/sensor precision.

Let us represent the second-order linearized system dynamics developed in the previous section in the following state-space form:

\begin{multline}
    \begin{bmatrix}
    \dot{{\eta}}_{2_\phi} \\ \ddot{{\eta}}_{2_\phi}
    \end{bmatrix}  = \begin{bmatrix}
    0 & I \\
    -\mathbf{M}_k^{-1}\mathbf{K}_k &-\mathbf{M}_k^{-1}\mathbf{D}_k
    \end{bmatrix} \begin{bmatrix}
    {\eta}_{2_\phi} \\ \dot{{\eta}}_{2_\phi}
    \end{bmatrix}   +\begin{bmatrix}
    0 \\ \mathbf{B}_{k_\gamma} \end{bmatrix} \tilde{\gamma} \\ +\begin{bmatrix}
    0  \\ \mathbf{B}_k
    \end{bmatrix} \tilde{w} + \begin{bmatrix}
    0 \\ \mathbf{B}_{k_\rho} \end{bmatrix} w_a,
    \label{eq:desc_class-k_dyn1}
\end{multline}

\noindent where $~~\mathbf{M}_k ~=~  V_{2_\phi}\T \Phi_2\T\mathcal{M} \Phi_2 V_{2_\phi},~~$ $\mathbf{D}_k ~=~  V_{2_\phi}\T \Phi_2\T\mathcal{D} \Phi_2 V_{2_\phi},~~$ $\mathbf{K}_k = V_{2_\phi}\T \Phi_2\T (\mathcal{K}+\mathcal{P} {K_s} (C_s \otimes I)) \Phi_2 V_{2_\phi}$, $\mathbf{B}_k = V_{2_\phi}\T \Phi_2\T\mathcal{P}$, and $\mathbf{B}_{k_\gamma} = - V_{2_\phi}\T \Phi_2\T\mathcal{P}{K_\gamma}$. 
Equation (\ref{eq:desc_class-k_dyn1}) represents the class-1 dynamics by applying the same idea to the equations developed in subsection \ref{ss:class1_min}.
In this problem formulation, we define the control input to be force density in the strings ($u = \tilde{\gamma}$), process noise to be ($w_p = \tilde{w} $), actuator noise to be $w_a$ (same coefficient matrix). Let us use the simple notations to denote the system described in equation~(\ref{eq:desc_class-k_dyn1}) as:

\begin{align}
\dot{x} &= A x +Bu + D_p w_p +D_aw_a,\\
y &= C_y x, \hspace{0.5 pc} \text{(output)}  \\ 
z &= C_zx+D_sw_s, \hspace{0.5 pc} \text{(measurement)} 
\end{align}
\noindent where $x = [{\eta}_{2_\phi}\T ~~ \dot{{\eta}}_{2_\phi}\T]\T $ is the state of the system and $u =  \tilde{\gamma} $ is the control vector.
The output of the system can be considered as the positions of the desired nodes $y = [n_i \T \cdots n_j \T]\T $ with appropriate definition of the system matrix $C_y$, and similar for the case of measurement vector $z = [n_i \T \cdots n_j \T]\T $. The vector $w_p, w_a$, and $w_s$ represent the disturbance noise, actuator noise and sensor noise, respectively. Let us consider the dynamic controller of the form:

\begin{align}
     \begin{bmatrix}{u}\\
\dot{{x}}_{ c}\end{bmatrix}=\left[\begin{array}{@{}c@{\quad}c@{}}{D}_c&{C}_c\\{B}_c&{A}_c\end{array}\right]\begin{bmatrix}{z}\\{x}_c\end{bmatrix}={G}\begin{bmatrix}{z}\\{x}_c\end{bmatrix},
\end{align}
with which the closed-loop system dynamics with  linearized system model and the above-mentioned dynamic controller can be written as:
\begin{align}%eqn1
\label{SYS}
\begin{bmatrix}
    \dot{x}_{cl}\\{y}\end{bmatrix}=\left[\begin{array}{@{}c@{\quad}c@{}}{A}_{cl}&{B}_{cl}\\{C}_{cl}&{D}_{cl} \end{array}\right]\begin{bmatrix}{x}_{cl}\\{w}\end{bmatrix}, 
\end{align}
% \begin{align}%eqn1
% \label{SYS}
% \begin{bmatrix}
%     \dot{{x}}\\{y}\\{z}\end{bmatrix}=\left[\begin{array}{@{}c@{\quad}c@{\quad}c@{}}{A}&{D}_p&{B}\\{C}_y&{D}_y&{B}_y\\C_z&{D}_z&{0} \end{array}\right]\begin{bmatrix}{x}\\{w}\\{u}\end{bmatrix}, 
% \end{align}
where $x_{cl} = [x^T~x_c^T]^T$, $w = [w_p^T~w_a^T~w_s^T]^T$ and all the system closed-loop matrices can be written in the standard form \cite{Skelton_LMI_1998}.

\begin{figure}[ht!]
\centering
\includegraphics[width=.95\linewidth]{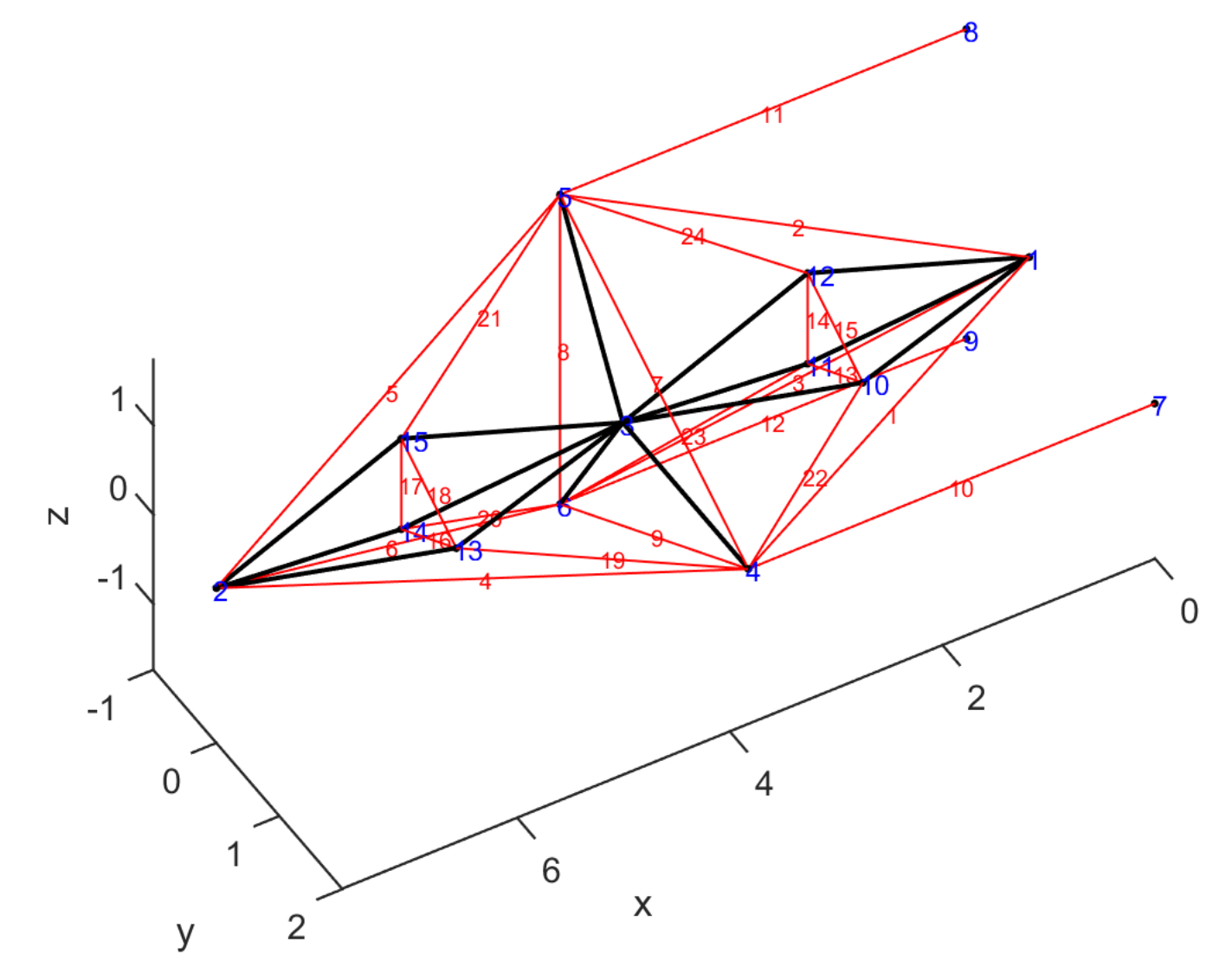}
\caption{Initial configuration of the tensegrity $T_1D_1$ robotic arm. The black lines represent the bars and red lines represent the strings. The string numbers are given in red and the nodes are numbered in blue.}
\label{f:T_1D_1_Lin}
\end{figure}

The example for all the different bounds will have the simulation results discussed on the same tensegrity $T_1D_1$ robotic arm with the initial configuration shown in Fig.~\ref{f:T_1D_1_Lin}. 
A {\tt MATLAB}\textregistered~based CVX toolbox is used for numerical implementation \cite{cvx}.
% Suppose that ${D}_y={0}$, ${B}_y={0}$, ${D}_p=[{D_{pn}} \ \ {D}_a \ \ 0]$, and ${D}_z=[{0} \ \ {0} \ \ {D}_s]$. 

% The vector ${w}^{ T} =\begin{bmatrix} {w}_p^{ T} & {w}_a^{ T} & {w}_s^{ T} \end{bmatrix}$ 
% contains process noise ${ w_p}$, actuator noise ${ w_a}$, and sensor noise ${ w_s}$. These disturbances are modeled as independent zero mean white noises with intensities ${ W_p}$, ${ W_a}$, and ${ W_s}$, respectively. The process noise intensity ${ W_p}$ was assumed to be known and fixed. The actuator and sensor precisions are defined as the inverse of the noise intensity (or variance in the discrete-time case) in each channel as:
% \begin{gather}
% \text{diag}({ \gamma_a}) \triangleq { \Gamma_a} \triangleq { W_a}^{-1}, ~~
% \text{diag}({ \gamma_s}) \triangleq { \Gamma_s} \triangleq { W_s}^{-1}.
% \end{gather}

%%%%%%%%%%%%%%%%%%%%%%%%%%%%%%%%%%%%%%%%%%%%%%%%%%%%%%%%%%%%%%%%%%%%%%%%%%%%%%%%%%%%%%%%%%%%%%%%%%%%%%%%%%%%%%%%%%%%%%%%%%%%%%%%%%%%%%%%%%%%%%%%%%%%%%%%%%%%%%%%%%%%%%%%%%%%%%

\subsection{Bound on Covariance in position error} 
It is impossible to drive the error to a precise zero in the presence of process noise; however, one can control the statistics of the error given the statistics of the noise. The covariance of the error in the position or velocity of the nodes can be bounded by bounding the covariance matrix $X$ for the linear system as \cite{Scherer_LMI_1997, Skelton_LMI_1998}:

\begin{align}
    \mathcal{E}[y y \T] = Y = CXC\T < \bar{Y},
    \hspace{1pc}
    \\
    A_{cl}X+XA_{cl}\T + B_{cl} W B_{cl}\T < 0,~X>0,
\end{align}
where the disturbance $w$ is modeled as independent zero mean white noises with intensities ${W}$.

\subsection{Bound on $\mathcal{L}_\infty$ norm of error or Generalized $\mathcal{H}_2$ Problem}
The peak value of a variable in the time domain is defined as $\mathcal{L}_\infty$ norm of the variable, i.e. $\|{y}\|_{\mathcal{L}_\infty}^2 = \sup [{ y}(t)\T y(t)]$. The following result provides a bound on peak value such that $\|{ y}\|_{\mathcal{L}_\infty}<\epsilon $, meaning that the peak value of
$[{ y}(t)\T{ y}(t)]$ is less than $\epsilon^2$ in the presence of finite energy disturbance. This problem can be solved as a ``energy to peak gain - $\Gamma_{ep}$" \cite{Skelton_LMI_1998} or generalized $\mathcal{H}_2$ problem \cite{Scherer_LMI_1997}.

\begin{align}
    \Gamma_{ep} &\triangleq \sup_{\|w\|_{\mathcal{L}_2} \leq 1} \|y\|_{\mathcal{L}_\infty},\\
    \Gamma_{ep} = \inf_Q \| CQC\T \|^{1/2}: & A_{cl}Q+QA_{cl}\T + B_{cl}B_{cl}\T < 0, ~ Q>0.
    \label{eq:Linf_gain}
\end{align}

\begin{figure}[ht!]
\centering
\includegraphics[width=1\linewidth]{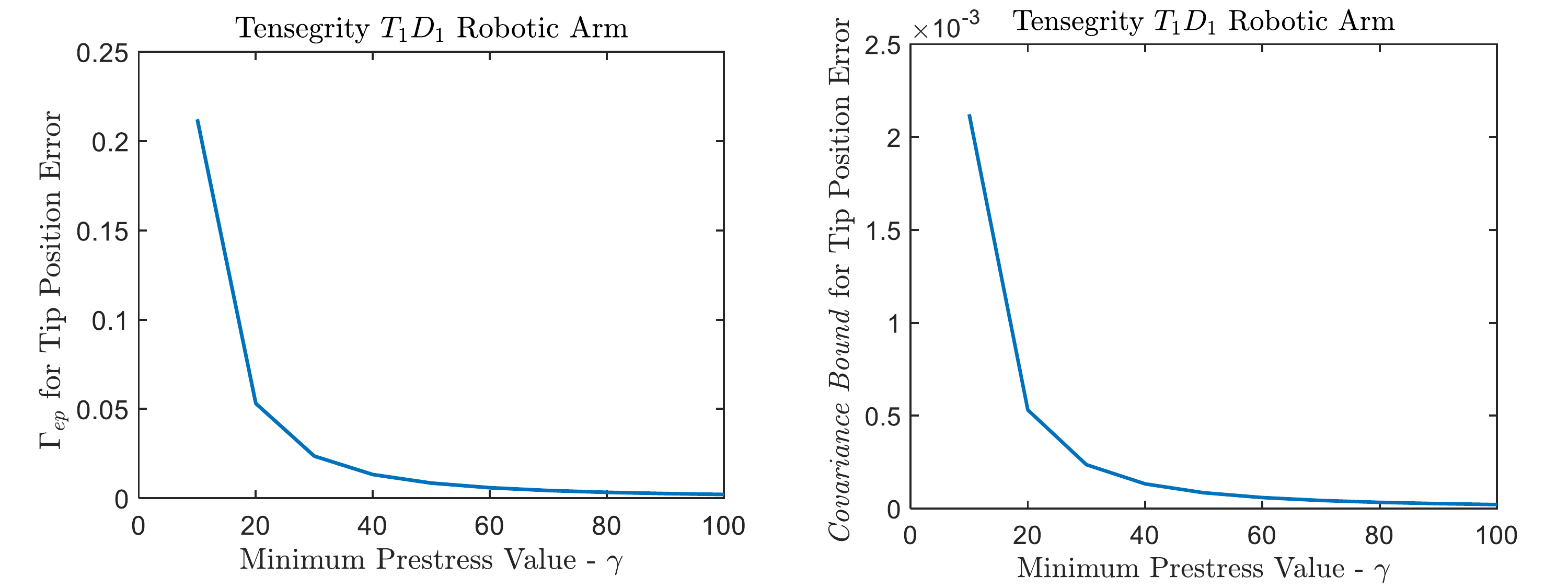}
\caption{Left: Bound on $\mathcal{L}_\infty$ norm of error for different values of the scaled prestress. Right: Bound the covariance error for different values of the scaled prestress.}
\label{f:Linear_OL_2_3}
\end{figure}

In Fig.~\ref{f:Linear_OL_2_3}, the plot on the left shows the bound on $\mathcal{L}_\infty$ norm of error for the tip for the unit energy disturbance applied to all the nodes of $T_1D_1$ robotic arm for different values of the scaled prestress in the strings. As the prestress value increases, the structure becomes stiffer which reduces the motion of the tip of the arm. The same trend can be observed for the plot on the right that the covariance bound on error for the tip of $T_1D_1$ robotic arm for different values of the scaled prestress in the strings.

\subsection{Bounded $\Gamma_{ie}$ or LQR Problem}
We define the linear quadratic regulator (LQR) problem to provide a performance bound $\epsilon > 0$ on the integral squared output such that $\|{y}\|_{\mathcal{L}_2} < \epsilon $ for any vector
${w}_0$ such that ${w}_0\T {w}_0 \leq 1$, and ${x}_0=0$. The disturbance $w$ is the impulsive disturbance ${w}(t) = { w}_0{\delta}(t)$. This can also be defined as the peak disturbance to energy gain ($\|{y}\|_{\mathcal{L}_2}$) for the system \cite{Skelton_LMI_1998}.

\begin{align}
\Gamma_{ie} &\triangleq \sup_{{w}_0{\delta}(t) \leq 1} \|y\|_{\mathcal{L}_2},\\
    \Gamma_{ie} = \inf_P \| B_{cl}\T P B_{cl} \|^{1/2}:& PA_{cl}+A_{cl}\T P + C\T C < 0, ~ P>0.
    % \label{eq:IE_gain}
\end{align}

\subsection{Bounded $\Gamma_{ee}$ or $\mathcal{H}_\infty$ Problem}
The result to bound the peak value of the frequency response of the transfer function $T(s) \triangleq  C(sI-A_{cl})^{-1}B_{cl}$ is now considered. The $\mathcal{H}_\infty$ Problem is defined as \cite{IWASAKI_Hinf_1994, Gahinet_Apkarian_1994}:

\begin{align}
     \|T\|_{\mathcal{H}_\infty} \triangleq \sup_w \|T(jw)\|< \epsilon
\end{align}
which can also be understood in time domain analysis as the energy-to-energy gain problem \cite{Skelton_LMI_1998}:

\begin{align}
    \Gamma_{ee} \triangleq & \sup_{\|w\|_{\mathcal{L}_2} \leq 1} \|y\|_{\mathcal{L}_2}< \epsilon, \\
    PA_{cl}+A_{cl}\T P + PB_{cl}&R^{-1}(PB_{cl})\T + C\T C < 0,\\  ~ P>0, ~ R &= \epsilon^2 I > 0.
\end{align} 

\begin{figure}[ht!]
\centering
\includegraphics[width=1\linewidth]{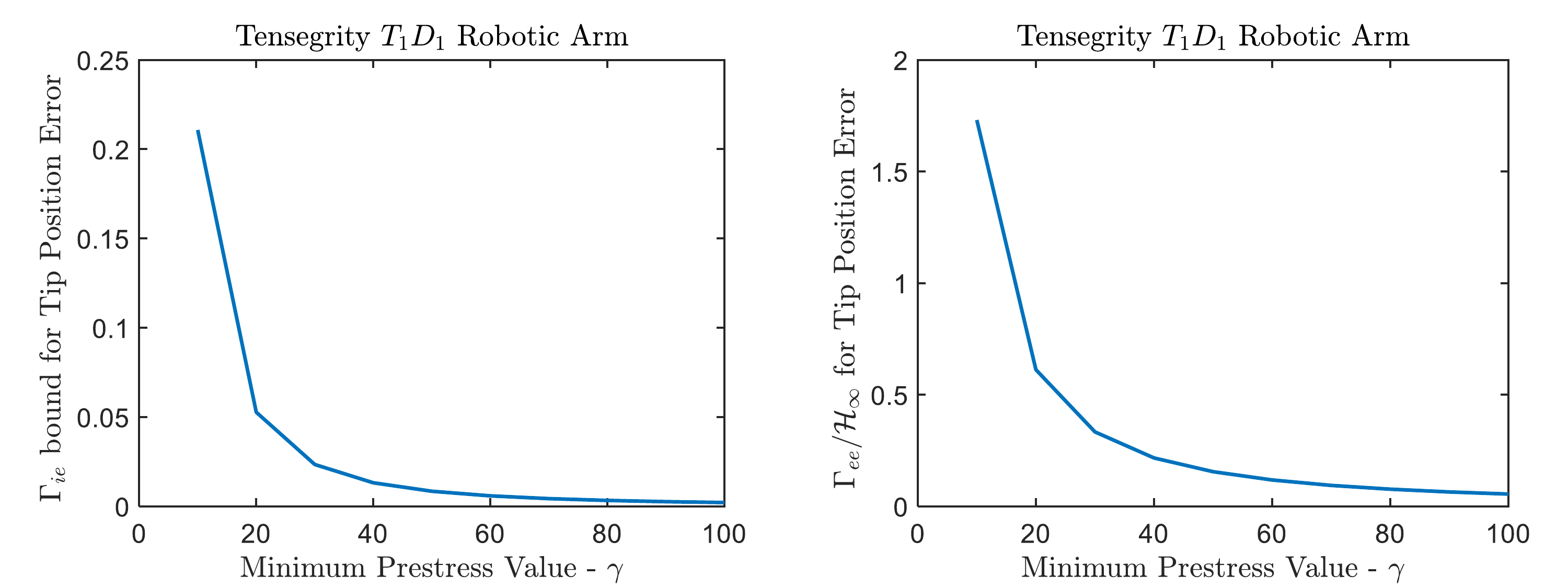}
\caption{Left: $\Gamma_{ie}$ bound for different values of the scaled prestress. Right: $\mathcal{H}_\infty$ bound for different values of the scaled prestress.}
\label{f:Linear_OL_4_5}
\end{figure}

The plots on the left in Fig.~\ref{f:Linear_OL_4_5} shows the impulse to energy bound for the error in the tip of $T_1D_1$ robotic arm for different values of the scaled prestress in the strings. The increased stiffness due to increased prestress reduces the values of $\Gamma_{ie}$. The same trend can be observed for the plot on the right for the $\mathcal{H}_\infty$ norm of the system which is also the gain from unit energy disturbance to energy in the error of the node position for the tip of $T_1D_1$ robotic arm. The value for the $\mathcal{H}_\infty$ norm decreases with the increased prestress.

%%%%%%%%%%%%%%%%%%%%%%%%%%%%%%%%%%%%%%%%%%%%%%%%%%%%%%%%%%%%%%%%%%%%%%%%%%%%%%%%%%%%%%%%%%%%%%%%%%%%%%%%%%%%%%%%%

% Figure~\ref{f:Arm_Hinf_2_Matlab} shows the response simulated using the optimal $\mathcal{H}_\infty$ controller in the closed-loop for the energy bounded disturbance applied to all the nodes, as shown in Fig.~\ref{f:Arm_H2_1}. No bounds on control input are added for the developed controller. The left plot shows the response with actuator and sensor noise of covariance $W_a = W_s = 10^{-5}$ and the right plot is one with zero sensor and actuator noise.

% \begin{figure}[ht!]
% \centering
% \includegraphics[width=1\linewidth]{figures/Arm_Hinf_2.pdf}
% \caption{Left: The closed-loop response with generalized $\mathcal{H}_\infty$ controller and actuator/sensor noise in the loop. Right: The closed-loop response with no actuator/sensor noise.}
% \label{f:Arm_Hinf_2}
% \end{figure}

% \begin{figure}[ht!]
% \centering
% \includegraphics[width=1\linewidth]{figures/Arm_Hinf_2_Matlab.pdf}
% \caption{Left: The closed-loop response with generalized $\mathcal{H}_\infty$ controller and actuator/sensor noise in the loop. Right: The closed-loop response with no actuator/sensor noise.}
% \label{f:Arm_Hinf_2_Matlab}
% \end{figure}

%%%%%%%%%%%%%%%%%%%%%%%%%%%%%%%%%%%%%%%%%%%%%%%%%%%%%%%%%%%%%
%%%%%%%%%%%%%%%%%%%%%%%%%%%%%%%%%%%%%%%%%%%%%%%%%%%%%%%%%%%%%

\section{Tensegrity System Dynamics in Descriptor Form}
Let us write the equation (\ref{eq:desc_class-k_dyn1}) for a class-k tensegrity structure in the following descriptor form:

\begin{multline}
    \begin{bmatrix}
    I & 0 \\
   0 & \mathbf{M}_k
    \end{bmatrix}\begin{bmatrix}
    \dot{{\eta}}_{2_\phi} \\ \ddot{{\eta}}_{2_\phi}
    \end{bmatrix}  = \begin{bmatrix}
    0 & I \\
    -\mathbf{K}_k &-\mathbf{D}_k
    \end{bmatrix} \begin{bmatrix}
    {\eta}_{2_\phi} \\ \dot{{\eta}}_{2_\phi}
    \end{bmatrix}    +\begin{bmatrix}
    0 \\ \mathbf{B}_{k_\gamma} \end{bmatrix} \tilde{\gamma} \\ +\begin{bmatrix}
    0  \\ \mathbf{B}_k
    \end{bmatrix} \tilde{w} + \begin{bmatrix}
    0 \\ \mathbf{B}_{k_\rho} \end{bmatrix} w_a.
    \label{eq:desc_class-k_dyn}
\end{multline}

% Equation (\ref{eq:desc_class-k_dyn}) can also represent class-1 dynamics when the subscript `$k$' is omitted from $\mathcal{M}_k,\mathcal{D}_k,\mathcal{K}_k,\mathcal{B}_k,\mathcal{F}_k$, and $\tilde{\eta}_2$ is replaced by $\tilde{n}$. 
In this problem formulation, we define the structure parameter to be initial prestress or force density at equilibrium condition ($\alpha = \bar{\gamma}$). It is to be noted that $\mathcal{K}_k(\bar{\gamma})$ is affine in initial prestress value $\bar{\gamma}$ as:

\begin{multline}
     \mathbf{K}_k(\bar{\gamma}) =    V_{2_\phi}\T \Phi_2\T \mathcal{T}\T P_{br}\hspace{0.2pc}(C_s\T  \otimes I) \widehat{(\bar{\gamma} \otimes \mathds{1})}(C_s \otimes I) \Phi_2 V_{2_\phi}  \\ + V_{2_\phi}\T \Phi_2\T \mathcal{T}\T K_{br}(\bar{f},\bar{b},\dot{\bar{b}})\mathcal{T}   \Phi_2 V_{2_\phi},
\end{multline}

\noindent where $K_{br}(\bar{f},\bar{b},\dot{\bar{b}})$ can also be written as some affine function of $\bar{\gamma}$ as $ K_{br}(\bar{f},\bar{b},\dot{\bar{b}}) = \mathcal{G} \bar{\gamma}$. Comparing it to system equation (\ref{state_eqn2}), we see only system matrix $A(\alpha)$ to be dependent on structure parameter.
% The other two output and measurement equations (\ref{meas_output_eqn2}) can be written to find the output as the position of certain nodes and to measure the position and velocity of certain nodes.

%%%%%%%%%%%%%%%%%%%%%%%%%%%%%%%%%%%%%%%%%%%%%%%%%%%%%%%%%%%%%%%%%%%%%%%%%%%%%%%%%%%%%%%%%%%%%%%%%%%%%%

%%%%%%%%%%%%%%%%%%%%%%%%%%%%%%%%%%%%%%%%%%%%%%%%%%%%%%%%%%%%%

\section{Tensegrity Example}
\begin{figure}[!htb]
    \centering
    \includegraphics[width=0.5\textwidth]{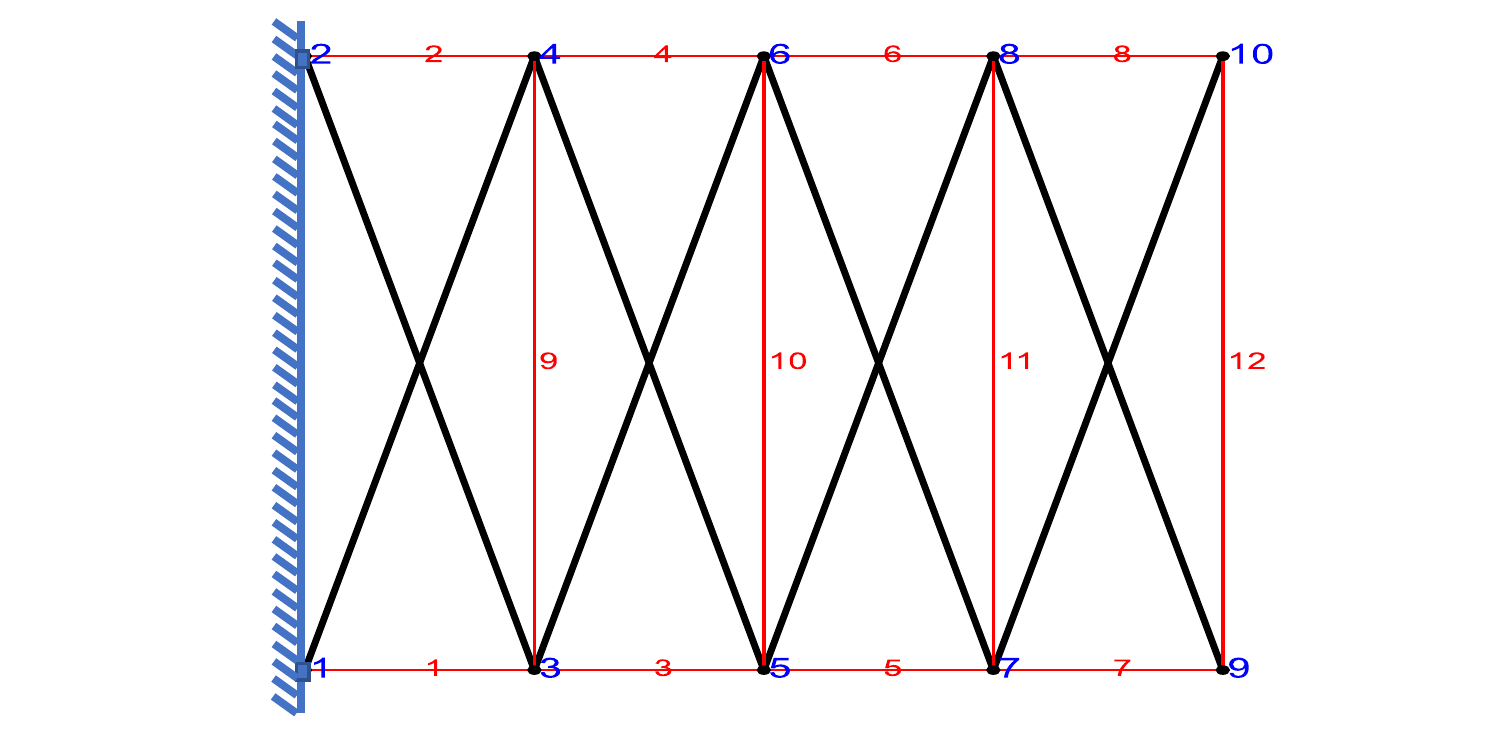}   
    \caption{2D Tensegrity Beam with 8 bars (shown in black) and 12 strings (shown in red). The node numbers are written in blue with string numbers in red.}
    \label{fig:Box-IASD-Fig}
\end{figure}

For a given 2D tensegrity structure shown in Figure \ref{fig:Box-IASD-Fig}, this section
optimizes the optimal prestresses in each string, precisions of sensors and actuators, and matrices corresponding to the dynamic controller for covariance control. The bars are shown in black, and the strings are shown in red. 
The mass for both the bars are assumed to be $m_b=1$kg and the mass for point mass is assumed to be $m_{s} = 0.5$kg. The tensegrity dynamics is linearized about the equilibrium configuration (corresponding to Figure \ref{fig:Box-IASD-Fig}) with minimum prestress values of $ \bar{\gamma} = 100$ and no external force.
\begin{figure*}[!ht]
\begin{multicols}{3}
    % \hspace{1.4cm}
    \centering
      \subfloat[Structure Parameter]{\includegraphics[width=1\linewidth]{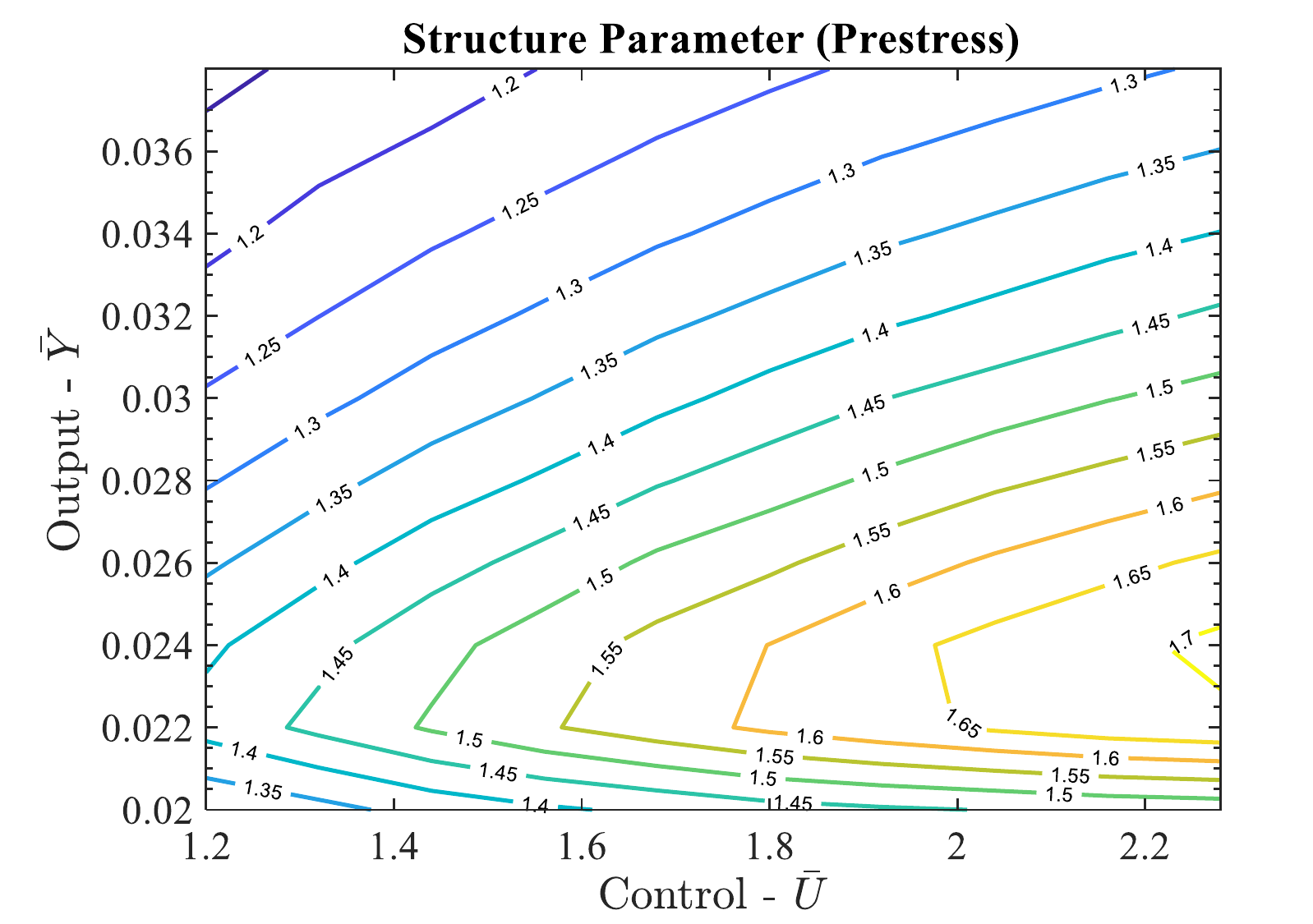}}
      \subfloat[Total Actuator Precision]{\includegraphics[width=1\linewidth]{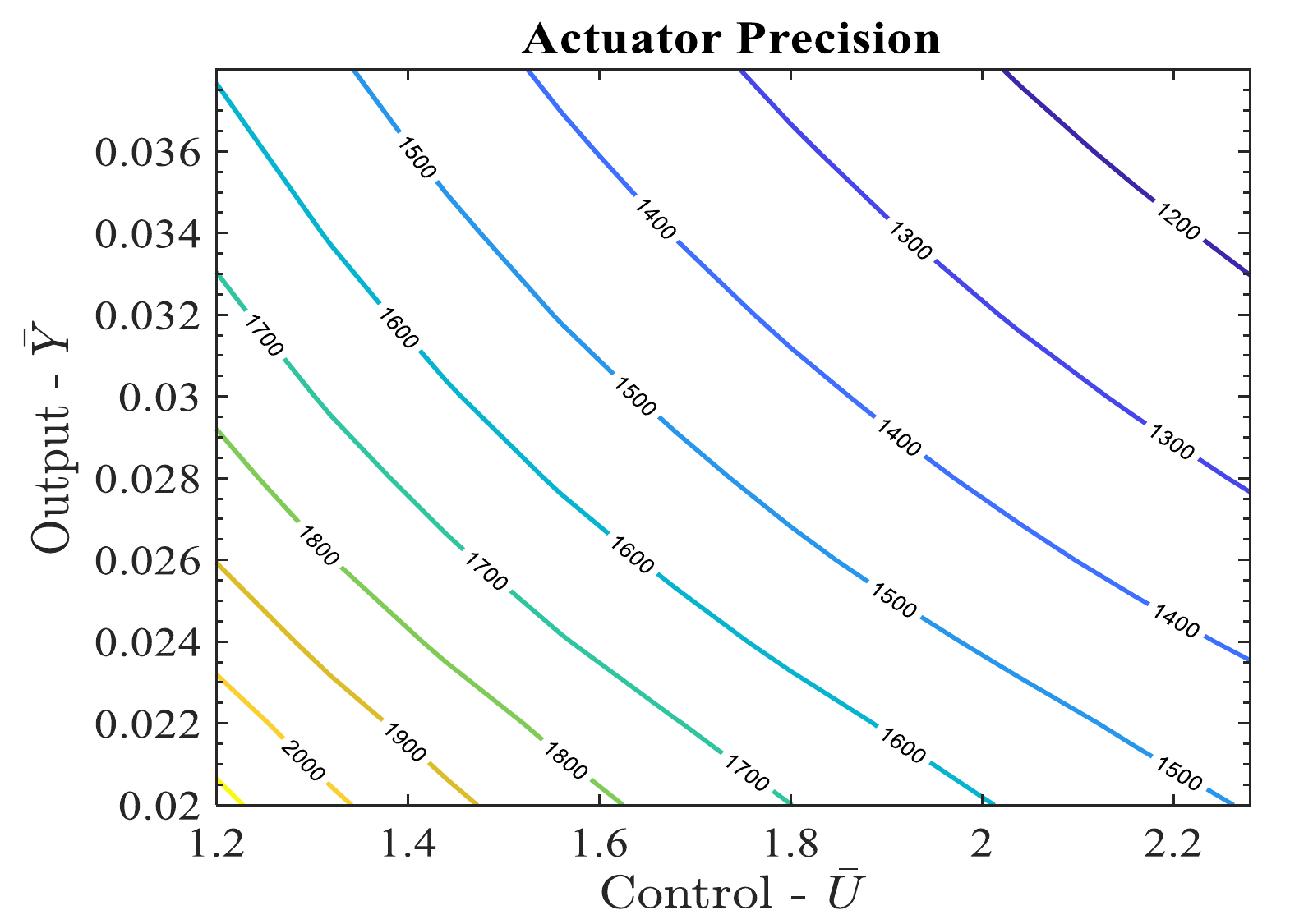}}
      \subfloat[Total Sensor Precision]{\includegraphics[width=1\linewidth]{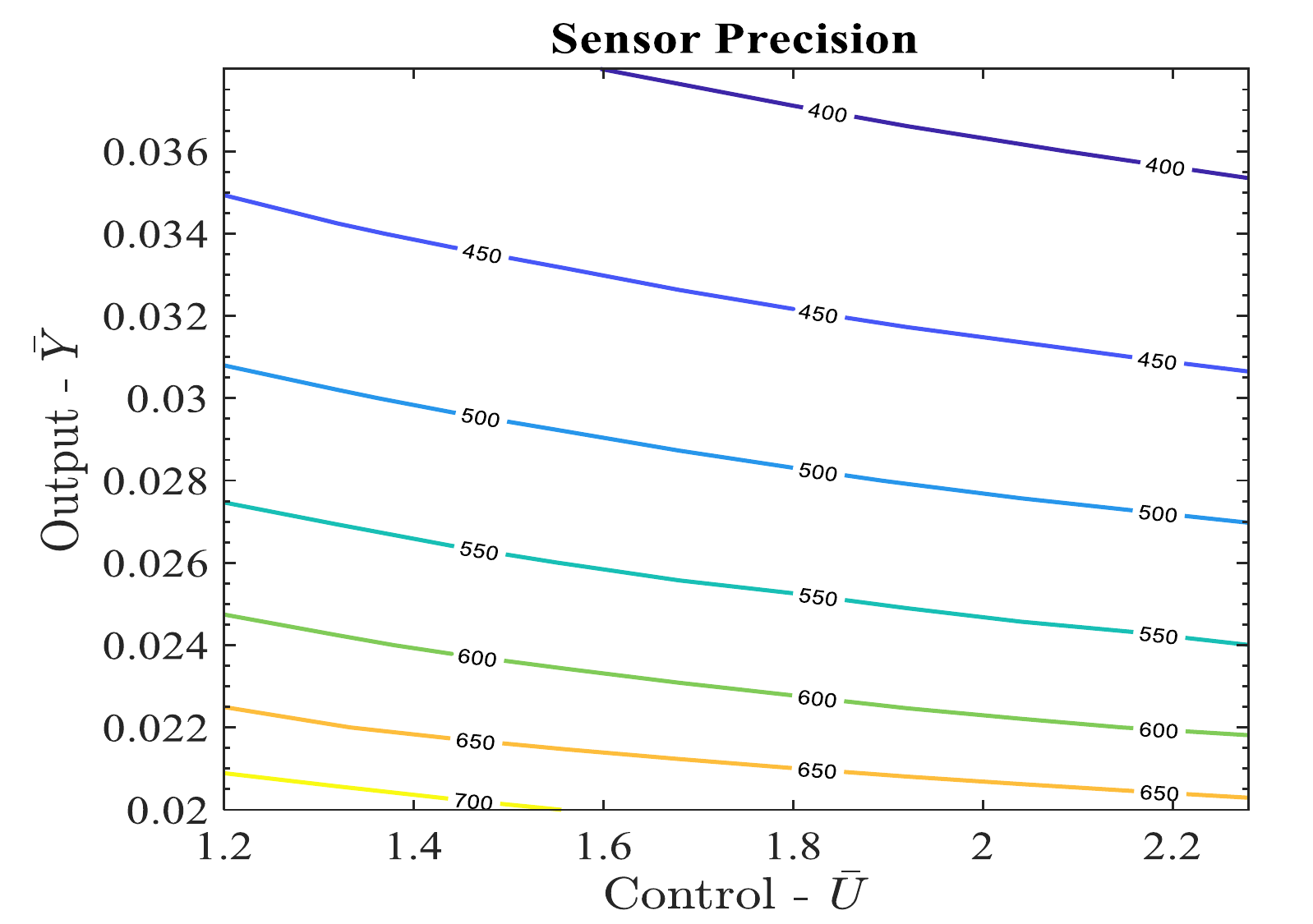}}
\end{multicols}
\caption{Plots with variation in input-$\bar{U}$ and output-$\bar{Y}$.}
\label{f:Box-IASD-Cntur}
\end{figure*}
The disturbances are modeled as external force hitting the nodes with a zero-mean white noise with intensity $W_p = 1$N$^2$. 
The disturbances are present on all the nodes with all strings as potential actuators.
The output to be bound is the top node at the far right of the beam (node 10) in x and y-direction. 
\begin{align}
    &y = \begin{bmatrix}
    n_{10x} \\ n_{10y} \end{bmatrix}  = C_y \begin{bmatrix}
    {\eta}_{2_\phi} \\ \dot{\eta}_{2_\phi}
    \end{bmatrix},~ C_y = \begin{bmatrix}
    0 & I \end{bmatrix} \Phi_2 V_{2_\phi} \begin{bmatrix}
    I & 0 \end{bmatrix},
\end{align}
and the measurements are the positions and velocity of all the nodes except nodes 1 and 2 as these nodes are fixed to ground.
\begin{align}
    z = C_z\begin{bmatrix}
    {\eta}_{2_\phi} \\ \dot{\eta}_{2_\phi}
    \end{bmatrix} + I w_s, ~ C_z = \begin{bmatrix} \Phi_2 V_{2_\phi} & 0 \\ 0 & \Phi_2 V_{2_\phi} \end{bmatrix}.
\end{align}
Some values assumed for this example are $\bar{\alpha}_{L} = 
0.1 \alpha_0$, $\bar{\alpha}_{U} = 10 \alpha_0,\bar{\gamma}_a = \bar{\gamma}_s = 1e3,p_a=p_s=1$, and $p_\alpha = 10.$

\begin{figure}[h]
    \centering
    \includegraphics[width=0.5\textwidth]{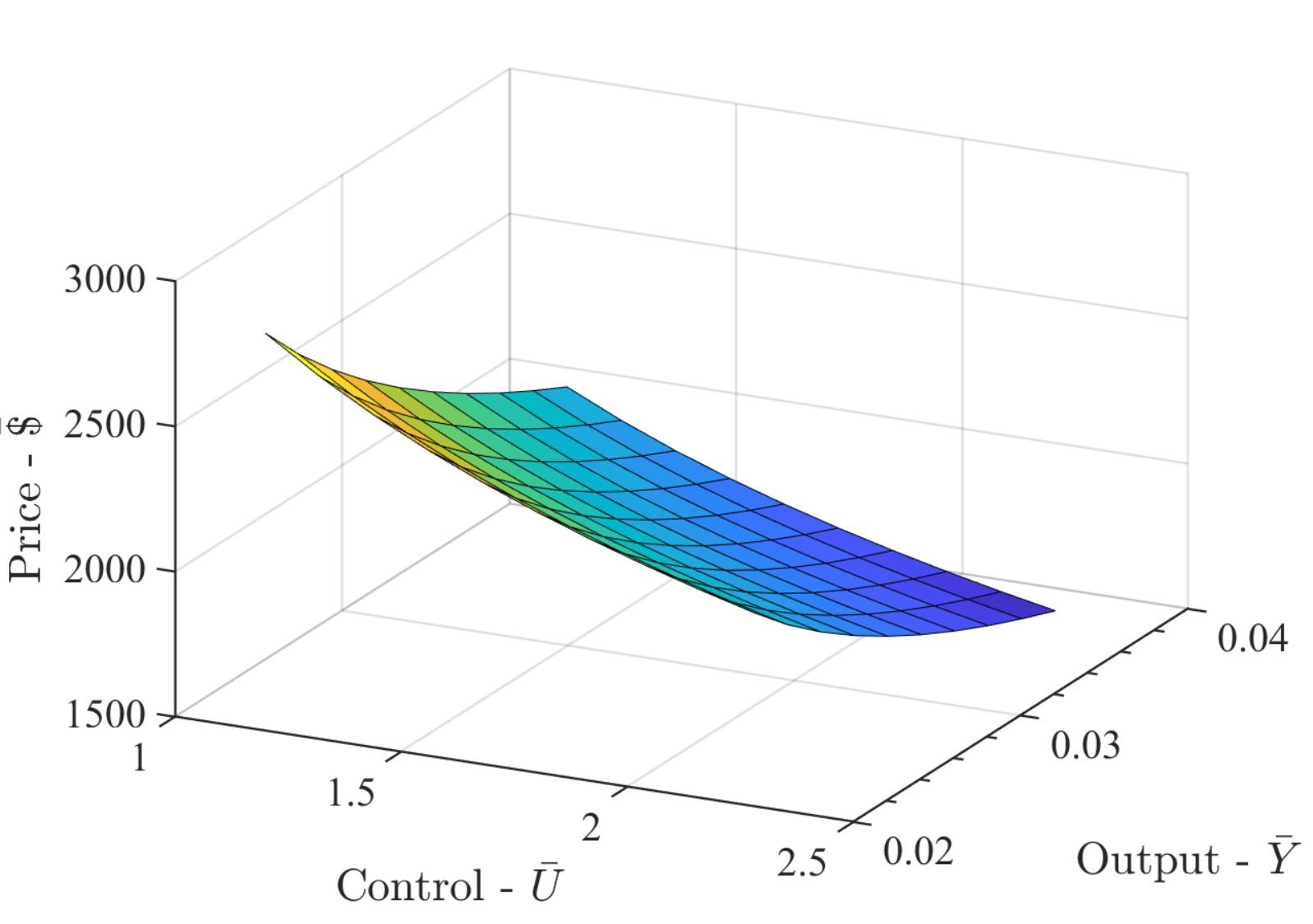}   
    \caption{The surface plot of the variation in budget with input-$\bar{U}$ and output-$\bar{Y}$.}
    \label{fig:Box-IASD-Surf}
\end{figure}

The surface plot in Figure \ref{fig:Box-IASD-Surf} shows the variation in budget requirement $\bar{\$}$ as we change the input covariance bound $\bar{U}$ and output covariance bound $\bar{Y}$. The required budget monotonically decreases with a more relaxed performance constraint for all values of control input bounds. The same decreasing trend in budget follows as we increase the control input bound. 
These trends can be understood as less precise sensors/actuators are required for relaxed constraint on the output bound or with the availability of higher control capability. 

% This small variation shows the relatively less strict nature of control covariance bound.

Figure \ref{f:Box-IASD-Cntur}(a) shows the contour curve for the prestress with variation in input-$\bar{U}$ and output-$\bar{Y}$. The figure shows the decreasing values of the prestress required for all strings as we increase the output covariance bound while maintaining the same control input $\bar{U}$, or to reduce the output covariance more prestress is required in the structure. Basically, less prestress is required for relaxed performance constraints.
The increase in prestress with relaxed input covariance can be understood due to less precision required on actuator and sensor, hence more money can be spent on pre-stress. 

% \begin{figure}[h]
%     \centering
%     \includegraphics[width=0.5\textwidth]{figures/Box-IASD-Cntur.pdf}   
%     \caption{The contour plot for the prestress value with variation in input-$\bar{U}$ and output-$\bar{Y}$.}
%     \label{fig:Box-IASD-Cntur}
% \end{figure}

Figures~\ref{f:Box-IASD-Cntur}(b) and \ref{f:Box-IASD-Cntur}(c) show the total actuator precision and sensor precision with variation in input-$\bar{U}$ and output-$\bar{Y}$. It can be observed that less precision on sensors and actuators is required for relaxed performance requirements along both the axes for output $\bar{Y}$ and input $\bar{U}$. This can be simply related to less budget requirement for the relaxed performance and control energy requirement from Fig.~\ref{fig:Box-IASD-Surf}.

% \begin{figure}[h]
%     \centering
%     \includegraphics[width=0.75\textwidth]{figures/Box-IASD-Cntur_ActSens.pdf}   
%     \caption{The contour plot for the total sensor precision and actuator precision with variation in input-$\bar{U}$ and output-$\bar{Y}$.}
%     \label{fig:Box-IASD-Cntur_ActSens}
% \end{figure}

\begin{table}[!h]
\caption{Sensor Precision for the measurement of position of $x$ and $y$ coordinates of all the nodes with $n_1$ and $n_2$ fixed to the wall (refer to Fig.\ref{fig:Box-IASD-Fig}).}
\centering
\begin{tabular}{|l|l|l|l|l|l|l|l|l|}
\hline
    & $n_3$    & $n_4$    & $n_5$   & $n_6$   & $n_7$   & $n_8$   & $n_9$   & $n_{10}$ \\\hline
x &  0.00 & 0.00 & 0.00 & 0.00 & 0.00 & 0.00 & 0.00 & 144.92\\\hline
y &  0.00 & 0.00 & 0.00 & 0.00 & 0.00 & 0.00 & 0.00 & 76.06  \\\hline
\end{tabular}
\label{T:IASD-sens_prec_pos}
\end{table}

\begin{table}[!h]
\caption{Sensor Precision for the measurement of velocity of $x$ and $y$ coordinates of all the nodes (refer to Fig.\ref{fig:Box-IASD-Fig}).}
\centering
\begin{tabular}{|l|l|l|l|l|l|l|l|l|}
\hline
& $n_3$ & $n_4$ & $n_5$ & $n_6$ & $n_7$ & $n_8$ & $n_9$ & $n_{10}$ \\ \hline
x &  5.49 & 12.75 & 8.75 & 17.47 & 12.31 & 46.59 & 13.61 & 164.17 \\ \hline
y & 0.00 & 0.00  & 0.00 & 0.00  & 0.00  & 0.00  & 73.24 & 155.39 \\ \hline
\end{tabular}
\label{T:IASD-sens_prec_vel}
\end{table}

\begin{table}[!h]
\caption{Actuator Precision for all the strings (refer to Fig.\ref{fig:Box-IASD-Fig}).}
\centering
\begin{tabular}{|l|l|l|l|l|l|l|}
\hline
Str No. & 1      & 2      & 3      & 4      & 5      & 6      \\ \hline
$\gamma_a$ & 168.22 & 173.29 & 113.03 & 203.72 & 182.25 & 240.43  \\ \hline \hline
Str No. &  7      & 8      & 9     & 10    & 11    & 12     \\ \hline
$\gamma_a$ &  104.17 & 597.00 & 24.46 & 43.92 & 74.53 & 200.21 \\ \hline
\end{tabular}
\label{T:IASD-act_prec}
\end{table}

The required sensor precision for output bound $\bar{Y} = 0.02$ and input bound $\bar{U} = 1.2$ is given in Tables~(\ref{T:IASD-sens_prec_pos} and \ref{T:IASD-sens_prec_vel}). The position sensors needs measurement only for the node that have output performance specified ${n}_{10x},{n}_{10y}$. However, for the velocity measurement, considerable precision is required to measure x-axis of all the nodes except the right most nodes and both x and y axis for the right most nodes,  $\dot{n}_{9x},\dot{n}_{9y}$,  $\dot{n}_{10x},\dot{n}_{10y}$.

Table~\ref{T:IASD-act_prec} gives the required precision values for all the actuators to achieve the performance bound of $\bar{Y} = 0.02$ with input constraint of $\bar{U} = 1.2$. The table shows higher precision is required on strings 8, and 12 as these strings are directly connected to the output node and the least precision is required for strings 9,10, and 11 as these strings do not directly affect the motion of the node $n_{10}$.

\begin{figure}[h]
    \centering
    \includegraphics[width=0.5\textwidth]{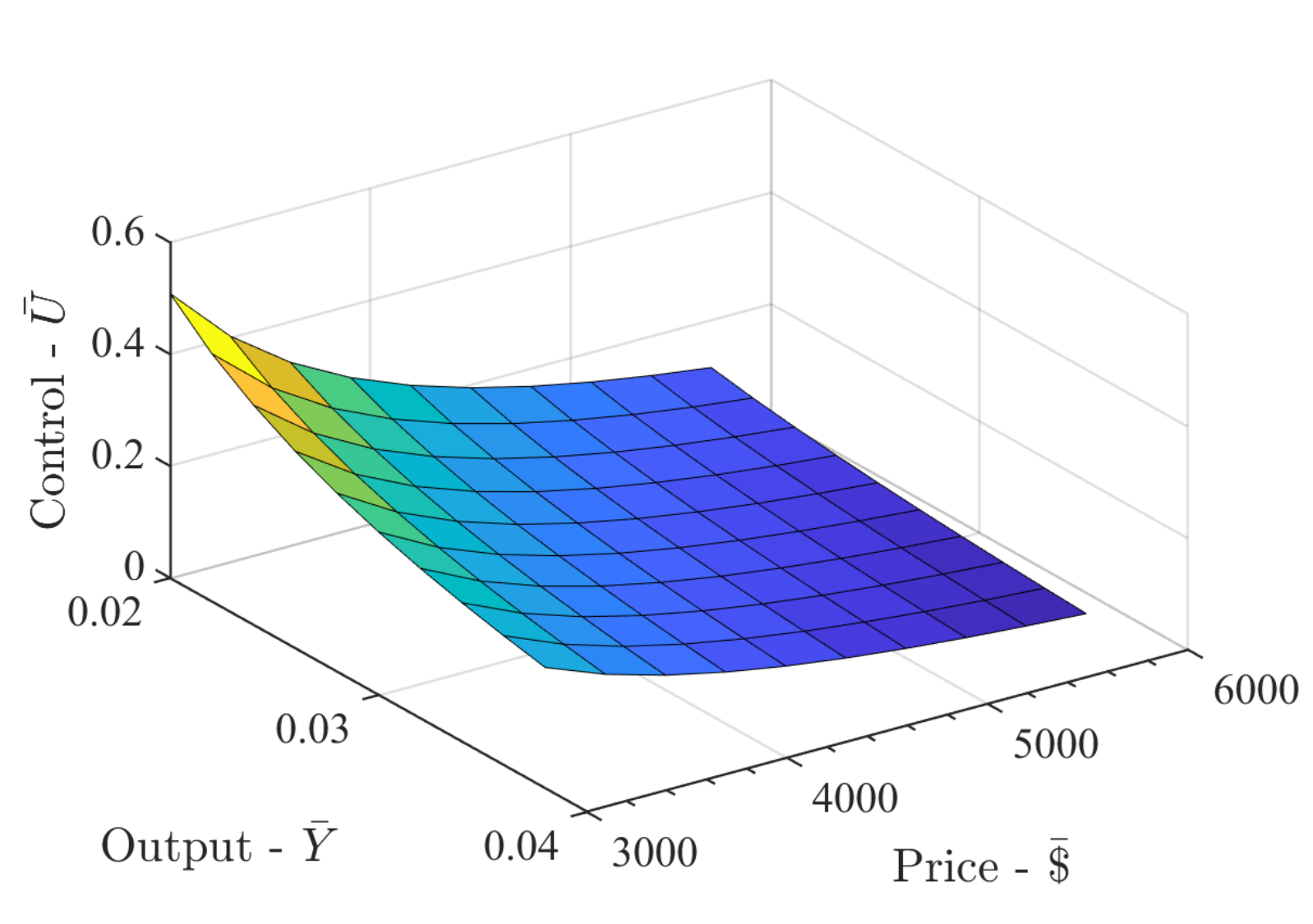}   
    \caption{The surface plot of the variation in input-$\bar{U}$ with budget $\bar{\$}$ and output-$\bar{Y}$.}
    \label{fig:Box-IASD-Ubar-Surf}
\end{figure}

\begin{figure*}[!ht]
\begin{multicols}{3}
    % \hspace{1.4cm}
    \centering
      \subfloat[Structure Parameter]{\includegraphics[width=1\linewidth]{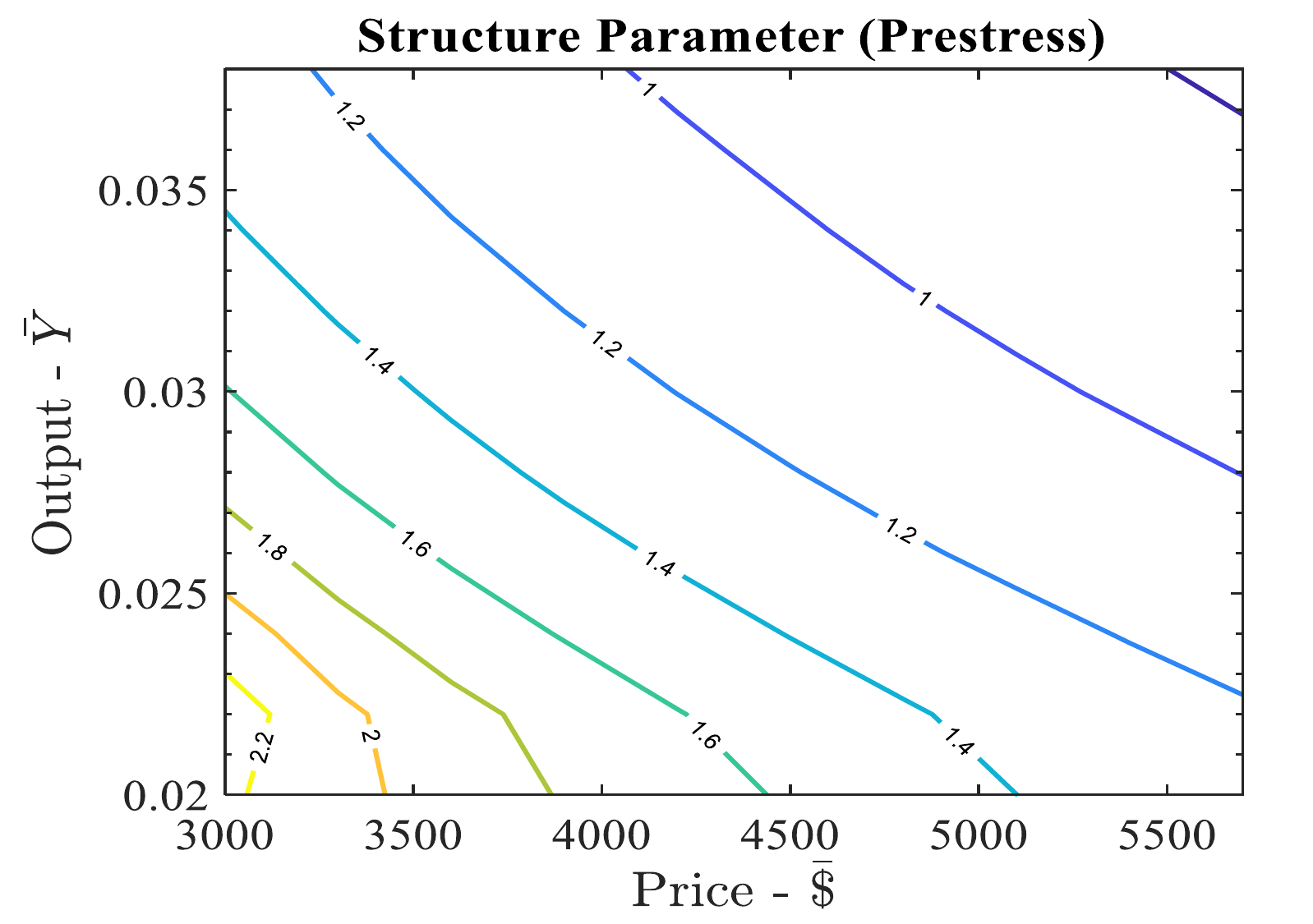}}
      \subfloat[Total Actuator Precision]{\includegraphics[width=1\linewidth]{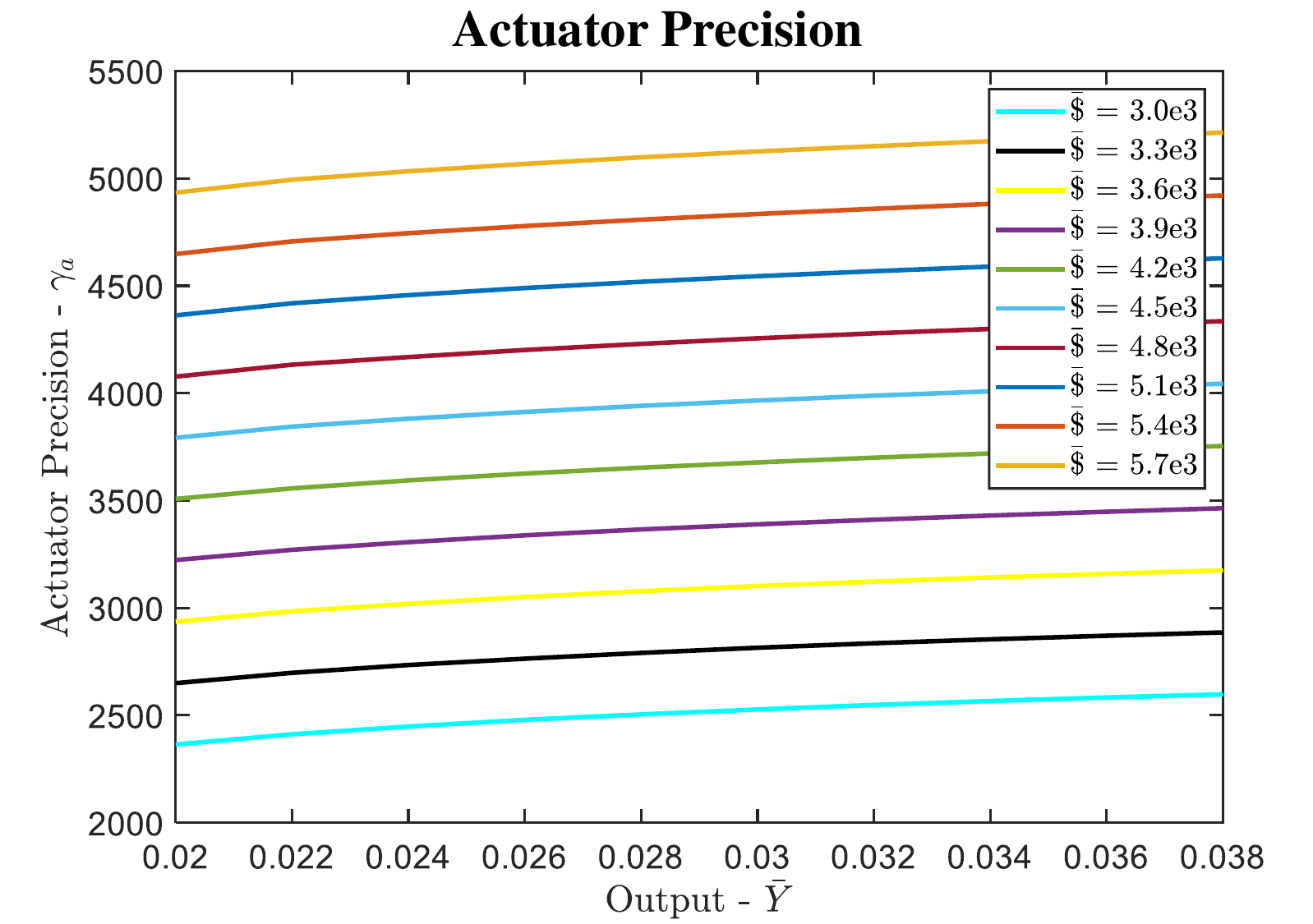}}
      \subfloat[Total Sensor Precision]{\includegraphics[width=1\linewidth]{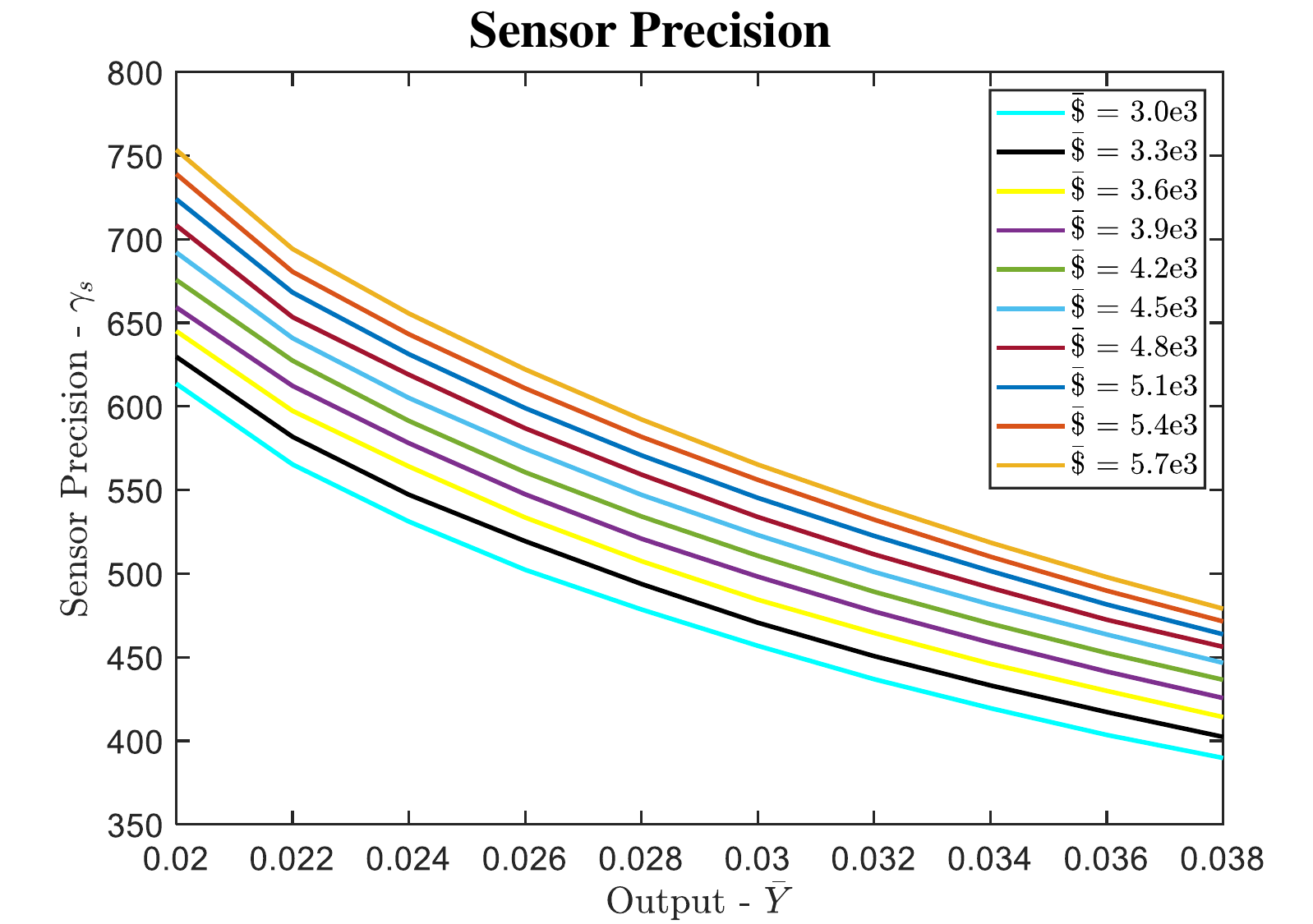}}
\end{multicols}
\caption{Plots with variation in Price-$\bar{\$}$ and output-$\bar{Y}$.}
\label{f:Box-IASD-Ubar}
\end{figure*}

A different trade-off is shown in Fig.~\ref{fig:Box-IASD-Ubar-Surf} and Fig.~\ref{f:Box-IASD-Ubar} by varying the bound on budget $\bar{\$}$ and output-$\bar{Y}$. Figure~\ref{fig:Box-IASD-Ubar-Surf} shows the surface plot variation in control input $\bar{U}$. The result shows that increasing the budget value $\bar{\$}$ and relaxing the output constraint $\bar{Y}$ will both result in smaller control input requirements as expected. Figure~\ref{f:Box-IASD-Ubar}(a) shows the contour plot for the prestress with the highest prestress required for the case of smallest budget-$\bar{\$}$ and tightest output bound-$\bar{Y}$. In Figs.~\ref{f:Box-IASD-Ubar}(b) and \ref{f:Box-IASD-Ubar}(c), we show several trajectories for different values of budget constraints as opposed to contour plots to show clear trends in the variation. The increase in the budget from $\$3000$ to $\$5700$ results in an increase in both sensor and actuator precision but notice that relaxing the output constraints results in an increase in actuator precision while we note a decrease in the value of sensor precision for the same budget value. This shows that the precision of the sensors is more important than the precision of the actuators to achieve smaller output covariance as the budget is being transferred from the actuators to the sensors.

% \begin{figure}[h]
%     \centering
%     \includegraphics[width=0.5\textwidth]{figures/Box-IASD-Ubar-Cntur.pdf}   
%     \caption{The contour plot for the total sensor precision and actuator precision with variation in input-$\bar{U}$ and output-$\bar{Y}$.}
%     \label{fig:Box-IASD-Ubar-Cntur}
% \end{figure}

% \begin{figure}[h]
%     \centering
%     \includegraphics[width=0.75\textwidth]{figures/Box-IASD-Ubar-ActSens.pdf}   
%     \caption{The contour plot for the total sensor precision and actuator precision with variation in input-$\bar{U}$ and output-$\bar{Y}$.}
%     \label{fig:Box-IASD-Ubar-ActSens}
% \end{figure}

%%%%%%%%%%%%%%%%%%%%%%%%%%%%%%%%%%%%%%%%%%%%%%%%%%%%%%%%%%%%%%%%%%%%%%%%%%%%%%%%%%%%%%%%%%%%%%%%%%%%%%

\section{Summary and Conclusion}
A novel system-level design approach by the simultaneous selection of control law, instrument precision, and structure parameters is developed. Nonlinear dynamics of the tensegrity system is linearized about an equilibrium position, and the modes corresponding to bar length change are removed to make the system physically realizable.
% The robust linear control theory was used to find the best performance bounds in terms of standard $\mathcal{H}_\infty$, generalized $\mathcal{H}_2$ formulation. Moreover, the precision of the actuators and sensors (information architecture) along with the controller is also solved for the covariance control problem. All the examples are shown on a $T_1D_1$ tensegrity robotic arm system. With the fixed structural parameters, the above-mentioned problem was proved to be convex having a globally optimal solution.
% The system dynamics is assumed to be linear along with the free structural parameters.
A dynamic controller is generated to bound the covariance of inputs and outputs with the precision of the sensors and actuators as the optimization variable. 
% The problem is set as a feasibility problem, where matrix upper bounds are specified for the covariance of selected outputs and the covariance of the control signals. 
% Specified upper bounds on the available precision of sensors/actuators and structural parameters are also given. 
The covariance control problem is formulated in an LMI framework where the combined optimization for structure parameter, sensor/actuator precision, and control law is shown to be a non-convex problem. The nonlinear matrix inequality constraints are approximated as linear matrix inequalities by adding a convexifying potential function. A sub-optimal solution is derived by iterating over the approximated convex problem.
The proposed system design approach is applied to a tensegrity system to design the structural system and information architecture simultaneously. 
% Results are provided for both dynamic compensation and full state feedback controller design.
% The tensegrity paradigm was used to integrate the structure and control design along with the information architecture. 
The linearized tensegrity dynamics model is used with initial prestress in the strings as a free structure parameter which appears linearly in the system matrices.
% The force density in the strings is used as the control input.
This paper allows us to design passive structures by reducing the level set of control covariance (to zero) while minimizing the output covariance constraint.
Trade-off analysis provided between cost vs. control energy and performance requirement showed that as performance and control energy constraints are relaxed, tighter budget constraints are achievable. 
% Tensegrity examples were used to show the effectiveness of the developed results. 
% The example results provide the knowledge on price estimate, what initial prestress to choose, where to put more precise sensors/actuators, and the controller parameters. 
Simulation results showcase a process where the price estimate, structure parameter (prestress) choice and sensor and actuator precision
trade-offs can be carried out simultaneously, leading to meaningful system design outcomes.

% \section*{Acknowledgment}

\ifCLASSOPTIONcaptionsoff
  \newpage
\fi

\bibliographystyle{IEEEtran}
\bibliography{IASD_refs,Raman_Tensegrity,Raman_Dynamics,Raman_Control}

\end{document}